\documentclass[preprint,aps]{revtex4}
\usepackage{graphicx}
\usepackage{booktabs}
\usepackage[usenames, dvipsnames]{color}
\usepackage{multirow}
\usepackage{eurosym}
\usepackage{setspace}
\usepackage{rotating}
\usepackage[ pdftex, plainpages = false, pdfpagelabels, 
                 pdfpagelayout = useoutlines,
                 bookmarks,
                 bookmarksopen = true,
                 bookmarksnumbered = true,
                 breaklinks = true,
                 linktocpage=all,
                 pagebackref=false,
                 colorlinks = true,
                 linkcolor = BrickRed,
                 urlcolor  = blue,
                 citecolor = BrickRed,
                 anchorcolor = green,
                 hyperindex = true,
                 hyperfigures
                 ]{hyperref}

\newcommand{\erf}{\mathrm{erf}}

\newcounter{figref}

\begin{document}

\title{\href{http://www.necsi.edu/research/economics/bondprices/}{The European debt crisis: Defaults and market equilibrium}} 
\date{\today}  
\author{Marco Lagi and \href{http://necsi.edu/faculty/bar-yam.html}{Yaneer Bar-Yam}}
\affiliation{\href{http://www.necsi.edu}{New England Complex Systems Institute} \\ 
238 Main St. S319 Cambridge MA 02142, USA}

\begin{abstract}
During the last two years, Europe has been facing a debt crisis, and Greece has been at its center. In response to the crisis, drastic actions have been taken, including the halving of Greek debt. Policy makers acted because interest rates for sovereign debt increased dramatically. High interest rates imply that default is likely due to economic conditions. High interest rates also increase the cost of borrowing and thus cause default to be likely. 
In equilibrium markets, economic conditions are used by the market participants to determine default risk and interest rates, and these statements are mutually compatible. If there is a departure from equilibrium, increasing interest rates may
contribute to---rather than be caused by---default risk.
Here we build a quantitative equilibrium model of sovereign default risk that, for the first time, is able to determine if markets are consistently set by economic conditions.
We show that over a period of more than ten years from 2001 to 2012, the annually-averaged long-term interest rates of Greek debt are quantitatively related to the ratio of debt to GDP. The relationship shows that the market consistently expects default to occur if the Greek debt reaches twice the GDP. 
Our analysis does not preclude non-equilibrium increases in interest rates over shorter timeframes. We find evidence of such non-equilibrium fluctuations in a separate analysis. According to the equilibrium model, the date by which a half-default must occur is March 2013, almost one year after the actual debt write-down. Any acceleration of default by non-equilibrium fluctuations is significant for national and international interventions. The need for austerity or other measures and bailout costs would be reduced if market regulations were implemented to increase market stability to prevent the short term interest rate increases that make country borrowing more difficult. We similarly evaluate the timing of projected defaults without interventions for Portugal, Ireland, Spain and Italy to be March 2013, April 2014, May 2014, and July 2016, respectively. The markets consistently assign a country specific debt to GDP ratio at which default is expected. All defaults are mitigated by planned interventions. 
\end{abstract}

\maketitle


\section{motivation}
Europe has been facing what the media has called ``a ferocious debt crisis'' since the beginning of 2010 \cite{Castle2011}. Several European governments have accumulated what many consider to be unsustainable levels of government debt \cite{Rastello2010}. Greece has been at the center of the crisis, with the highest levels of public debt in the Eurozone and one of the biggest budget deficits \cite{eurostat}. Fears of default and its consequences \cite{Korowicz2012} have led to special bailout funds \cite{Nelson2011}, austerity programs in Greece \cite{OGrady2010}, and recently agreements by banks to voluntarily dismiss half of Greece's debt \cite{Fidler2011}. Policy makers have taken action because of the dramatic increase in interest rates, which curtails the ability to borrow money from international capital markets. High interest rates imply that the market considers the risk of default to be high due to economic conditions, but they also increase the cost of borrowing and cause default to be more likely. Because interest rates both reflect expected default risk and cause default risk, it is not entirely clear which is the actual cause of default: economic conditions, i.e. an equilibrium market, or non-equilibrium market behavior.

In equilibrium economics, market prices are based upon value and value is determined by price. This self-consistency in equilibrium hides the important subtle causal loops of the interaction of price with value that become manifest when deviations from equilibrium occur. 
Among the causes of deviations from equilibrium are herd behaviors that cause price changes whose timing is not related to inherent changes in value, but only to the intrinsic dynamics of the market itself. Such deviations from equilibrium may play a role in fluctuations characterized by ``volatility'' (though volatility can also include variations in equilibrium prices and value) but the impact of fluctuations away from equilibrium is poorly understood because they depend on those poorly characterized causal loops.

In recent years, major market indices have varied widely, amid concerns about macroeconomic conditions that affect broadly the values of goods and services. In the case of markets whose values depends on macroeconomic conditions, the equilibrium market price reflects economic news. When market price is itself a causal factor in economic conditions, which it often is, the feedback to economic conditions can lead to inherent dynamics that are not found in the theory of equilibrium prices. Because markets are considered to represent economic conditions, the interpretation of market changes is used in policy setting for economic interventions. When non-equilibrium fluctuations of the market are interpreted as indicators of economic conditions for policy interventions, this misinformation can lead to policy actions that are not justified by economic conditions but only by the false signals given by market fluctuations. However, in some markets, prices directly impact on the economic system. This is the case in bond markets, where interest rates that are set by the market directly impact on the borrowing costs of countries. Whether the markets are or are not in equilibrium is therefore key to both economic conditions and correct policy response. Disentangling the perception of market fluctuations from economic conditions, and the possibility of market fluctuations determining rather than reflecting economic conditions is important to both economic theory and policy. While equilibrium prices are important components of the proper and constructive function of economic markets, the impact of non-equilibrium fluctuations may be destructive, and therefore should be the subject of policy attention, which can by careful choice limit both the size and impact of non-equilibrium volatility.

Here we build a simple equilibrium model of sovereign default risk that, for the first time, is able to directly quantify the consistency between interest rates and economic indicators, and serve as a basis for evaluating the role of non-equilibrium volatility in the bond market.
The first answer we reach is that the equilibrium picture holds for an analysis of annually-averaged long-term interest rates. Looking more closely in a second analysis \cite{Lagi2012}, we show that this picture breaks down for timeframes less than a year long during which non-equilibrium speculator behavior plays a key role, leading to large interest rate fluctuations.
In particular, bond prices dropped from 57\% to 21\% of their face value, from July to December 2011.
A complex systems perspective suggests that near a transition large fluctuations between distinct states (in this case default and non-default) become possible. Under these conditions interest rate variations are both caused by, and cause, vulnerability to default. Such fluctuations can trigger substantial economic impacts and stronger interventions than would be justified by equilibrium markets. The extent of these fluctuations may be reduced by well chosen market regulations.


\section{overview}
It is common practice for private corporations and governments to borrow in order to overcome shortfalls in operating expenses, deal with emergencies, or invest in economic growth. A primary mechanism for this borrowing is to sell interest bearing bonds. The question of how much governments can borrow from bond markets is related to the lenders' concern about debt repayment and sustainability \cite{Eaton1981,Cooper1984}. The levels of government debt vary markedly from country to country, being the result of a mixture of policy decisions (if levels are low) and how much lenders will finance (if levels are high). The interest rates are set by the willingness of those who loan funds, either by direct negotiation or by auction. Only if the issuer becomes unable to repay, i.e. a default happens, are the payments reduced. The interest on the primary bond market therefore is, in first approximation, a measure of how high the risk of default is. The more likely those who loan the funds think default is, the higher interest rates are set, reflecting a greater investment risk.
Different bonds provide more or less interest depending on their estimated relative risk of default.

Many researchers have addressed the question of why nations ever choose to pay off their debts, given sovereign immunity and no enforcing body to exact repayment \cite{Alfaro2005, Cole1998, Cooper1984, Panizza2009, Arraiz2006, Lindert1989, Eaton1981, Diaz-Alejandro1983, Mitchener2005, Tomz2007book, DePaoli2006, Wright2011, Dooley2000, Sandleris2008, Cohen1992, Alichi2008, Perotti1996, Borensztein2009, Sturzenegger2007b}. Most assume that there is some cost associated with defaulting, which may include loss of reputation in the international community \cite{Alfaro2005, Cole1998, Panizza2009}, trade sanctions \cite{Diaz-Alejandro1983, Wright2011}, harsher future credit terms \cite{Arraiz2006, Borensztein2009, Sturzenegger2007a}, or outright exclusion from the world credit market \cite{Arraiz2006, Eaton1981}. Historically, military interventions may have served as an additional external deterrent \cite{Mitchener2005,Panizza2009}. However, some question the importance of such external repercussions, claiming that they may not be as stringently enforced or as effective as traditionally thought \cite{Lindert1989, Tomz2007book, Panizza2009}. By whatever mechanism, it is clear 
from measures of economic growth that sovereign default correlates with subsequent reduced economic performance of the defaulting country \cite{DePaoli2006, Wright2011, Dooley2000, Sturzenegger2004, Panizza2009}.

A complementary body of literature attempts to identify the warning signs that precede sovereign default \cite{Aguiar2006, Arellano2007, Hernandez-Trillo2000, Kulatilaka1987, Manasse2005, Savona2008, Stein2001, Tomz2007}. Researchers have cited macroeconomic measures of insolvency and illiquidity as precursors to default; they consider GDP growth, debt-service payments, penalties for default, bond interest rates, interest volatility, consumption volatility, and a host of other factors \cite{Manasse2005, Kulatilaka1987, Savona2008, Arellano2007}. Some have validated their models on empirical data \cite{Arellano2007, Hernandez-Trillo2000, Manasse2005, Savona2008, Yue2010}. However, many of them include a large number of parameters, obscuring the relative importance of different factors.

Here we develop a quantitative model of sovereign default that identifies the debt ratio \cite{Domar1944} as the only relevant dynamic macroeconomic variable that European market participants have used over the last 10 years to estimate the likelihood of default of a country and, therefore, to set long-term interest rates. The model explicitly relates the annually-averaged long-term interest rates to the debt ratio with a simple 2-parameter fit, and is able to accomplish two major goals.

\begin{figure}[tb]
\refstepcounter{figref}\label{fig:equilibrium}
\href{http://www.necsi.edu/research/economics/bondprices/equilibrium.pdf}{\includegraphics[width=0.7\linewidth]{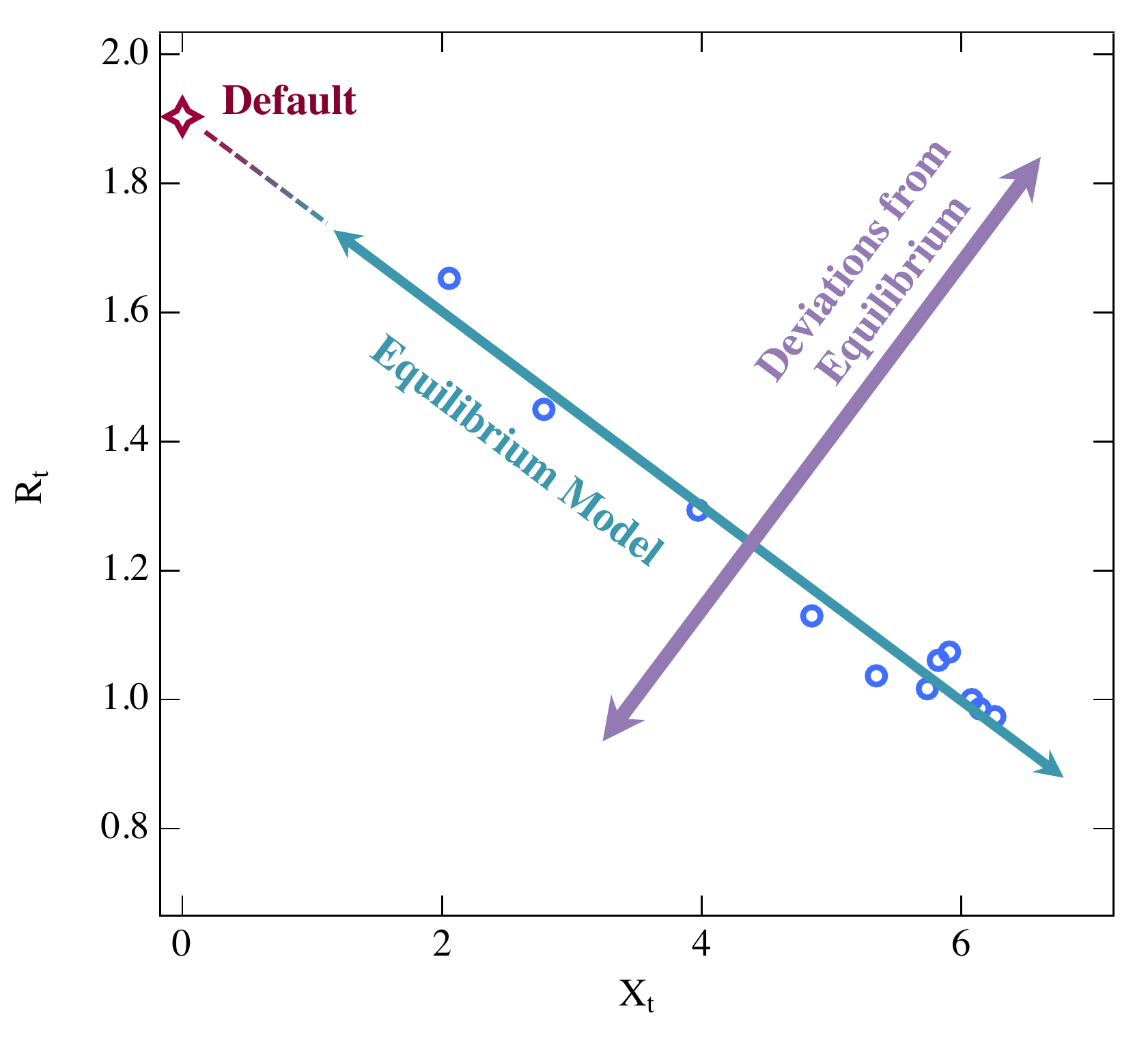}}
\caption{\textbf{Test of equilibrium market model of default} - Greece's debt ratio $R_t$ (blue dots) as a function of the parameter $X_t$, which reflects annually-averaged long-term interest rates. The equilibrium model derives a linear relationship between $R_t$ and $X_t$ (see Eq. \ref{eq:linear_intext}). All points, from 2001 to 2011, lie on a straight line, indicating that over this period of time markets have consistently determined a level of debt at which default occurs---the debt for $X_t=0$ (maroon star).}
\end{figure}

First, the equilibrium model provides a way of testing whether international capital markets are consistently basing interest rates on economic conditions.
We show that interest rates reflect the expectation that default becomes certain as the debt to GDP ratio approaches a specific value, the default threshold. 
Although there is strong evidence for the existence of mechanisms able to drive prices away from equilibrium \cite{Litterick2002, food_prices}, economic models usually assume that markets are reflecting economic information rather than considering it a hypothesis to be tested. 
We test the assumptions of the equilibrium picture by showing that the annually-averaged long-term interest rates for Greek bonds over the last 10 years are given by a well-defined function of the distance of the sovereign debt ratio from a fixed country-specific default threshold. 
Interest rates follow this function toward default as the debt ratio increases (see Fig. \ref{fig:equilibrium}). Since the interest rates can be quantified as an explicit function of macroeconomic information, the model shows that annually-averaged long-term interest rates are consistent with this assumption of equilibrium models.

\begin{figure}[tb]
\refstepcounter{figref}\label{fig:extrapolation}
\href{http://www.necsi.edu/research/economics/bondprices/gr_extrapolation_final.pdf}{\includegraphics[width=0.7\linewidth]{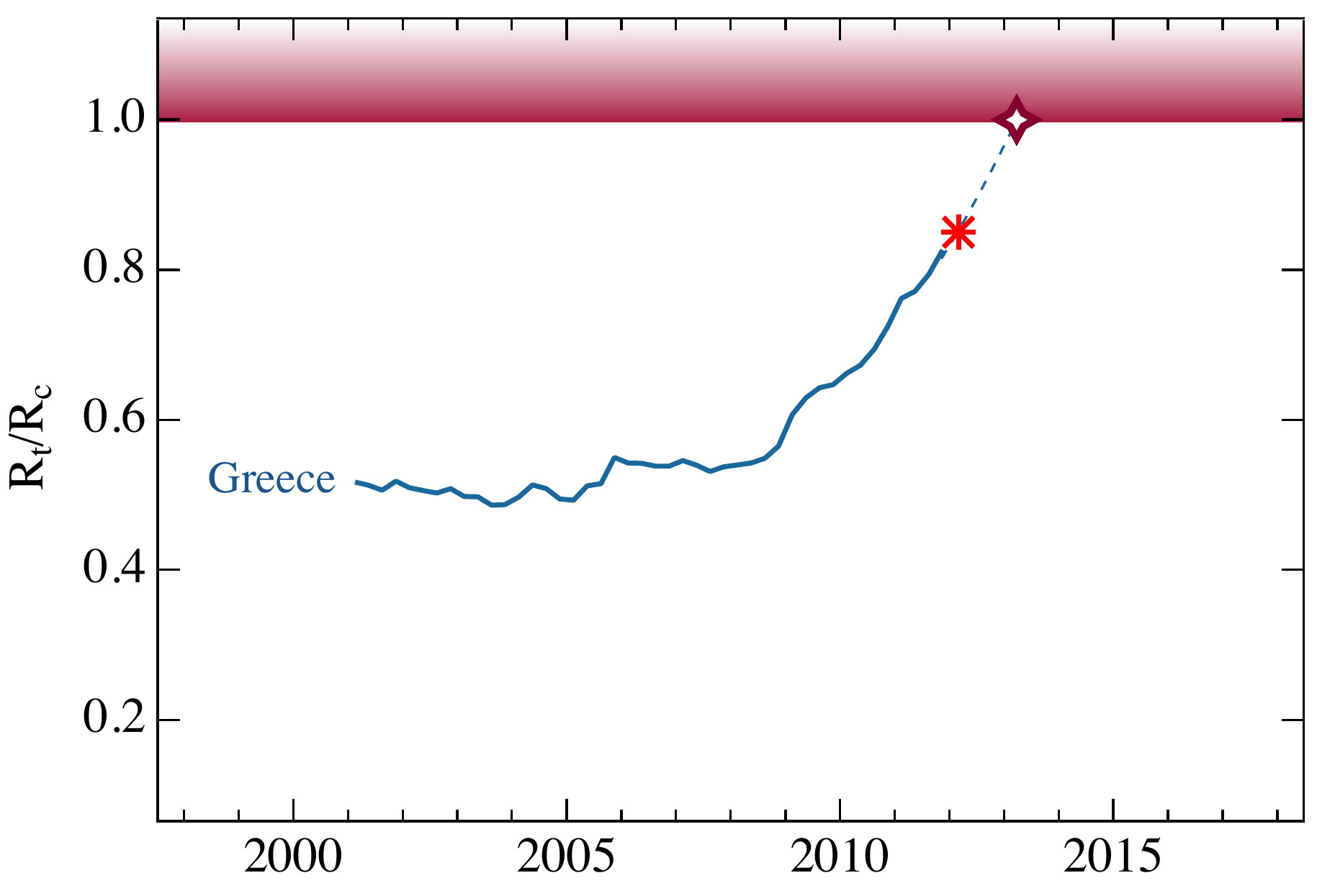}}
\href{http://www.necsi.edu/research/economics/bondprices/extrapolation_final.pdf}{\includegraphics[width=0.7\linewidth]{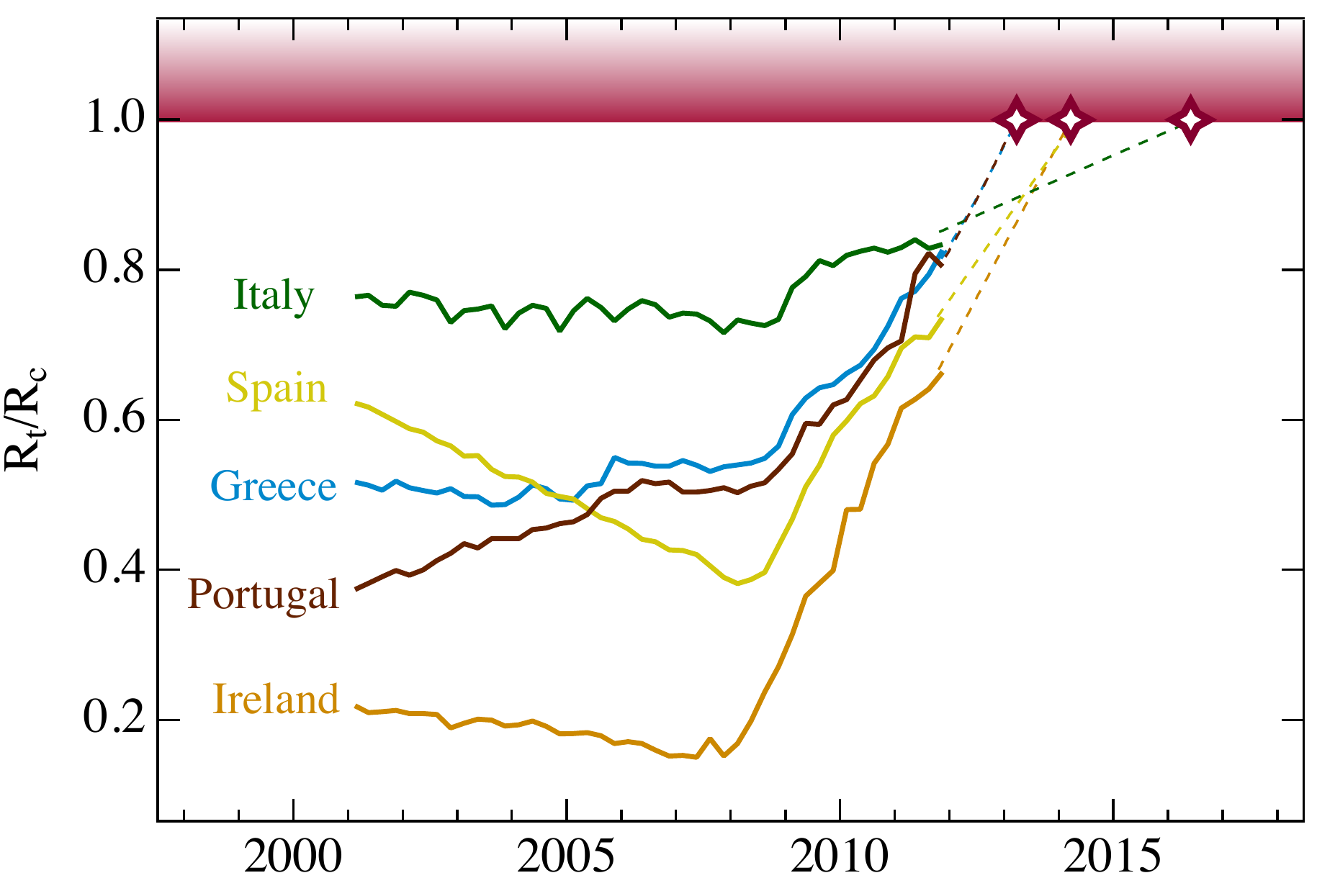}}
\caption{\textbf{Timing of default} - \emph{Top Panel}: Greece's debt ratio $R_t$ normalized by our model parameter $R_c$, the critical debt ratio, as a function of time (solid line). The projected debt trajectory (dashed line), intersects the normalized threshold in March 2013. Red mark indicates the actual partial default in March 2012. \emph{Bottom Panel}: Similar analysis for all five EU countries considered in this work: Italy, Spain, Greece, Portugal and Ireland. See Appendix \hyperref[app:a]{A} for more details on the model, and Appendix \hyperref[app:d]{D} for more details on individual countries.}
\end{figure}

Second, the model can estimate the timing of a default event given its debt trajectory. 
Default becomes certain as the country debt approaches the debt threshold. 
Even when a country defaults a portion of the debt is typically honored, the amount of which is called the \emph{recovery rate}, affecting the precise timing of the default. 
For Greece, we find that the model points to certain default with a 50\% recovery rate in the first quarter of 2013 (Fig. \ref{fig:extrapolation}, top). The ``voluntary'' write-down of half of the country's debt by private creditors, which in effect corresponds to a partial default with 50\% recovery, occurred one year earlier in March 2012 \cite{Gatopoulos2012}. The timing of default may be attributed to non-equilibrium market fluctuations in the secondary bond market and not to economic or other news. Thus, the equilibrium trajectory of default is preempted by market bandwagon effects. Results for other countries are summarized in the bottom panel of Fig.\ \ref{fig:extrapolation}. The projected timing of default ranges from March 2013 for Portugal to July 2016 for Italy.

Our analysis shows, therefore, that the change of long-term interest rates follows fundamental expectations only over the long run, while it is subject to non-equilibrium effects in the short run. We discuss this in a second paper, where we show that shorter time variation in bond prices have been much more rapid than would be expected from economic fundamentals \cite{Lagi2012}. The results are consistent with the existence of price fluctuations due to trend following or market manipulation. These fluctuations are both caused by and cause vulnerability to default. When there is a concern about default, large fluctuations are created by herd behavior. When interest rates fluctuate upwards they can cause default. Under these conditions the causal relationships are reversed and markets drive fundamental economic outcomes. These effects shorten the time to default, causing a need for more aggressive austerity programs and debt write-downs than might otherwise be required. During a period of economic recovery from a recession these differences may be critical as a recovering economy may be better able to avoid default if it is not subjected to higher national interest rates and austerity measures than would be justified in equilibrium markets. 
 
In considering the broader implications of our analysis of country default, it is important to note that our model was developed and validated in the context of the European debt crisis, which is an atypical context due to the common euro
currency that prevents individual country currency devaluation for members of the Eurozone. Where currency devaluation is possible, devaluation may serve to reduce some of the pressure of paying back internal debt, also by making exports cheaper and generating growth \cite{Deo2011}. Thus, the European conditions may have distinctive properties and an analysis of European conditions may not be applicable elsewhere without modification.

In particular, one of our central findings is that only one macroeconomic variable is dynamically relevant, the debt ratio. The debt ratio has been recognized to be a common reference for lenders among measures that are expected to be important \cite{Cooper1984}. Our model allows for the possibility that other variables play a role in country default as long as they either do not change over time, and therefore contribute only to the static model parameters, or are not relevant in the specific context of the Eurozone countries we studied. Indeed, other variables must play a role in determining the value of the static country specific debt threshold. Our analysis does not reveal these variables. Still, the debt thresholds are in a limited range between 90\% and 200\% of GDP for the countries we studied---implying markets consider the GDP itself to be a first approximation to a reasonable debt threshold.

This conclusion may, however, be modified for other countries outside the Eurozone. A variable that may not be relevant in the Eurozone because of the common currency, but could be relevant in other areas, is the distinction between external and internal debt \cite{Buchanan1957}. Where one class of lender is substantially less likely to stop providing loans than others, an analysis may be best framed in terms of the distinct properties of the two levels of available debt. It is reasonable to expect that the internal debt plays a diminished role in the probability of default. If the ratio between the internal and external debt were constant over time, discounting the role of the internal debt in default would increase the value of the critical debt ratio, but would not otherwise affect our conclusions.

Thus, we can expect our findings to be modified for countries where internal debt is a large part of total debt, especially for countries outside of the Eurozone. For example, Japan has a high debt ratio of 230\% but much of the debt is internal and owned by the postal system \cite{Oxford2009}. Much of the US debt is owned by the Federal Reserve \cite{Toscano2012}. In both these cases, interest rates are low despite debt ratios that are in the range of the Eurozone debt thresholds.

Finally, the applicability of the model is limited to countries in distress. We cannot know what the critical threshold of the debt ratio is when interest rates are low: the only way to test the model is when the probability of default is high enough to affect the annually-averaged long-term interest rates.
 
 
 \section{model}
We construct a mathematical model of sovereign default risk (details are in Appendix \hyperref[app:a]{A}) by considering a country to have the choice, at each time step, of being in a state of \emph{default} or \emph{nondefault}. Since the ability of a country to pay its debt depends on the size of the debt relative to its economic output, we assume that the choice of default is determined by its \emph{debt ratio}, i.e. the ratio between the overall debt and the GDP. We also assume the existence of a critical threshold $R_c$ of the debt ratio, above which the country is likely to default. The value of the critical threshold depends on the cost of the penalties implicitly associated with defaulting; it differs from country to country and is the first parameter of our model.

\begin{figure}[tb]
\refstepcounter{figref}\label{fig:feedback}
\href{http://www.necsi.edu/research/economics/bondprices/feedback.pdf}{\includegraphics[width=0.8\linewidth]{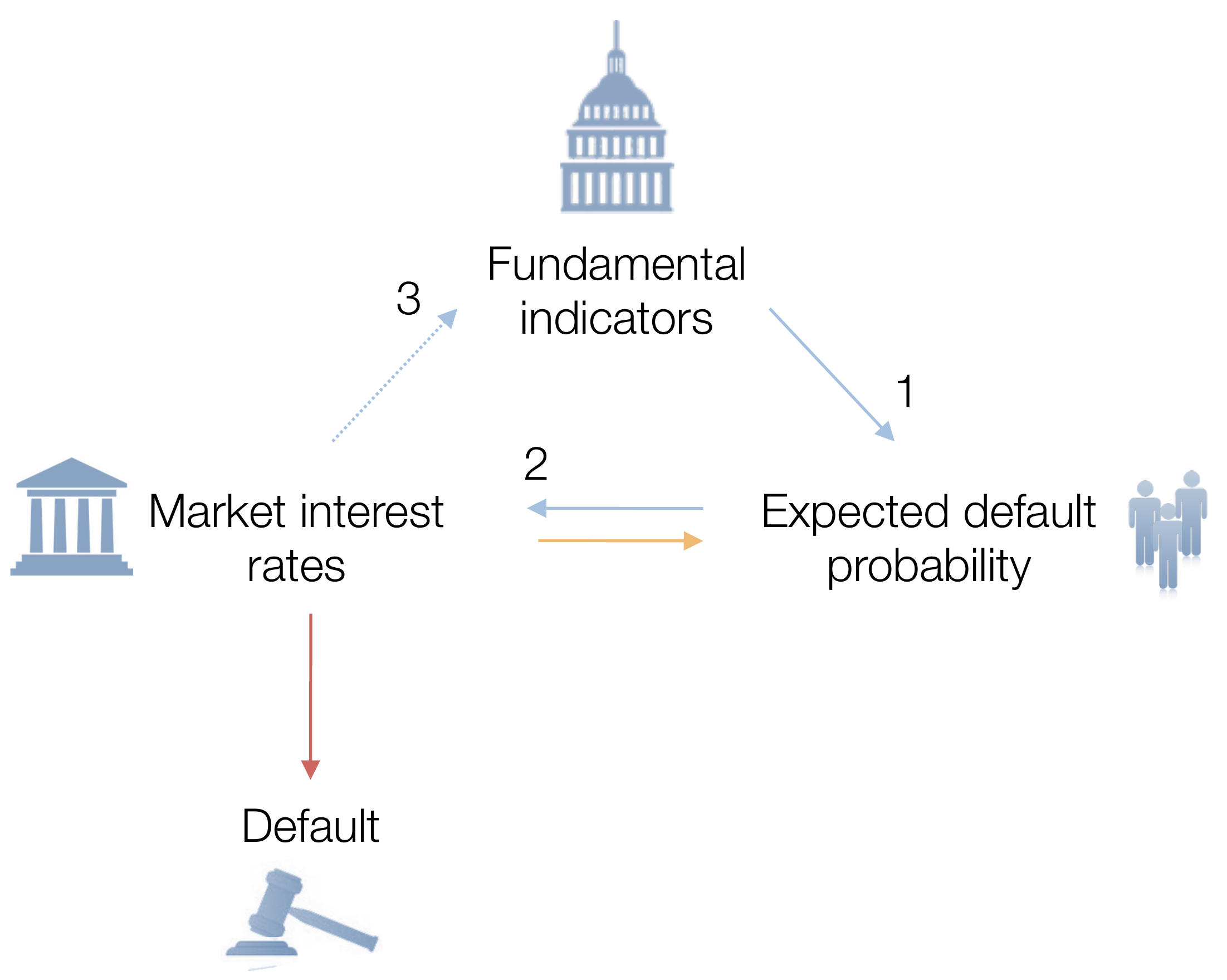}}
\caption{\textbf{Interest rate setting mechanism} - In equilibrium, investors use macroeconomic indicators of national production and debt to estimate the probability of default (blue arrow marked 1) using Eq. \ref{eq:final_p_text}.  The market then sets bond prices and interest rates according to that risk (blue arrow marked 2) using Eq. \ref{eq:pdefault_bond}. Interest paid on the debt contributes to the subsequent debt (blue arrow marked 3), according to Eq. \ref{eq:ratio}). The model can also determine whether sudden changes in interest rates driven by bandwagon effects increase the perception of default probability (orange arrow), pushing interest rates out of equilibrium.}
\end{figure}

The expected probability of default $P_t$ is indicated by market participants through the setting of interest rates: the more likely they think default is across all possible future scenarios over the period of repayment, the higher interest rates are set, reflecting a greater investment risk. Since interest rates determine the debt burden, their increase also determines a higher probability of default. This positive feedback mechanism is shown as a blue dashed arrow in Fig. \ref{fig:feedback}. Moreover, a sharp change in interest rates can alter investor perception of the default probability (orange arrow), thereby creating another feedback loop and pushing interest rates out of equilibrium. There are therefore two distinct feedback mechanisms, one fundamental and one behavioral. The model can test whether the behavioral feedback has an impact on the markets, or whether markets are self-consistently determining a default probability based on the fundamental feedback (blue arrows).

If all lenders had the same expectation about the time of default given by a default threshold, the interest rate $i_t$ would be the risk free value until the debt ratio $R_t$ 
reaches the critical default value. At this point, creditors would stop lending money and interest rates would effectively diverge. The resulting step-like profile for the probability of default is in practice smooth, since their expectations are heterogeneous. Many reasons contribute to this heterogeneity, including the limited amount of information available to the economic actors in the bond market, different estimates of the critical debt ratio, imperfect knowledge of the debt trajectory, different evaluations of the political decision-making process behind a possible default, different bond maturation periods and payment schedules, and disparate influence of other economic indicators. Heterogeneous expectations have been previously recognized to be important in describing the behavior of bond markets \cite{Xiong2010}. Nevertheless, the probability of default $P_t$ is low below $R_c$ and high above, so that the overall profile as a function of the debt ratio $R_t$ can be represented by a smooth sigmoidal transition,

\begin{equation}
\displaystyle
P_t = \left[1-\frac{1}{1+e^{(R_t-R_c)/\eta}}\right]\frac{1}{1-\rho}
\label{eq:final_p_text}
\end{equation}
where $\eta$, the second parameter of our model, reflects the degree of smoothness, and $\rho$ is the extrinsically set recovery rate, i.e. the probability that the country will pay a share of the debt $\rho < 1$. The derivation of this equation can be found in Appendix \hyperref[app:a]{A}.

\begin{figure}[tb]
\refstepcounter{figref}\label{fig:Pfits}
\href{http://www.necsi.edu/research/economics/bondprices/sig_fit.pdf}{\includegraphics[width=0.9\linewidth]{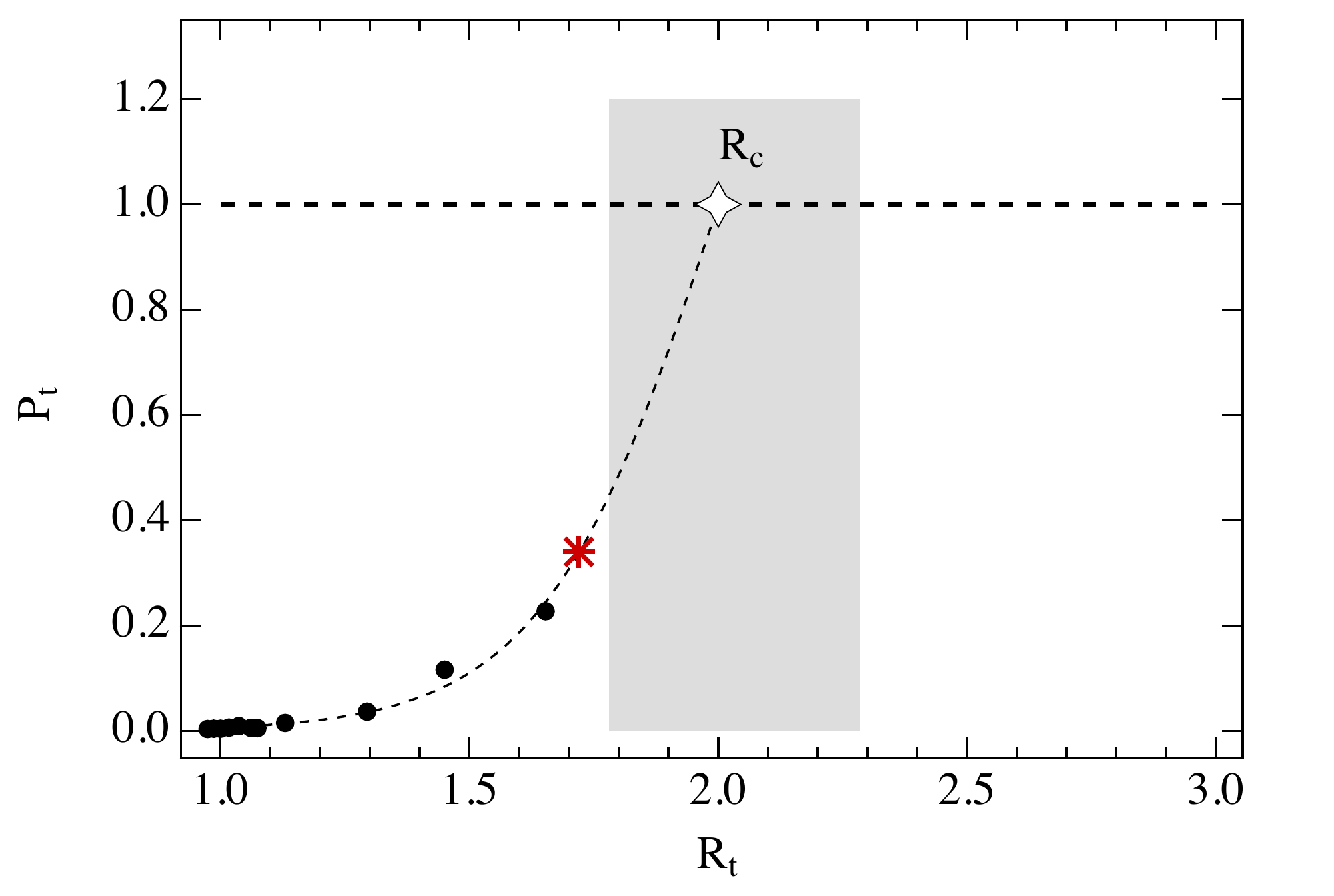}}
\caption{\textbf{Default model validation: Greece} -  Plot of the increasing probability of default with increasing debt according to data for Greece  (dots) and theory (dashed line, Eq. \ref{eq:final_p_text}) using the relationship of interest rates to risk as specified by Eq. \ref{eq:pdefault_text}. The red star indicates the actual partial default in March 2012 \cite{Gatopoulos2012}, at a time when the probability of default was $P_t=0.33$. The fit of theory to data yields the two model parameters, the critical debt ratio $R_c=2.00\pm0.07$ and the heterogeneity parameter $\eta = 0.18\pm0.02$. The data and theory shown are for $50\%$ recovery. Vertical shading shows the range of debt ratios of certain default for a range of recovery rates between 20\% to 80\%. German interest rates are used as risk-free interest rates.
Using quarterly data as opposed to annual data would introduce seasonality effects, but would not change significantly the results.}
\end{figure}

In summary, we consider the default process as one that follows a discrete jump at a particular value of the debt ratio, but we treat it as smoothed by uncertainty and heterogeneity. The model has only two parameters: the critical debt ratio $R_c$, which represents the average value above which investors stop buying bonds, and the heterogeneity parameter $\eta$, which incorporates market uncertainty. The model assumes that these parameters are well defined and consistently represented by the bond market over time.

\begin{figure}[tb]
\refstepcounter{figref}\label{fig:poly_fit}
\href{http://www.necsi.edu/research/economics/bondprices/poly_fit.pdf}{\includegraphics[width=0.9\linewidth]{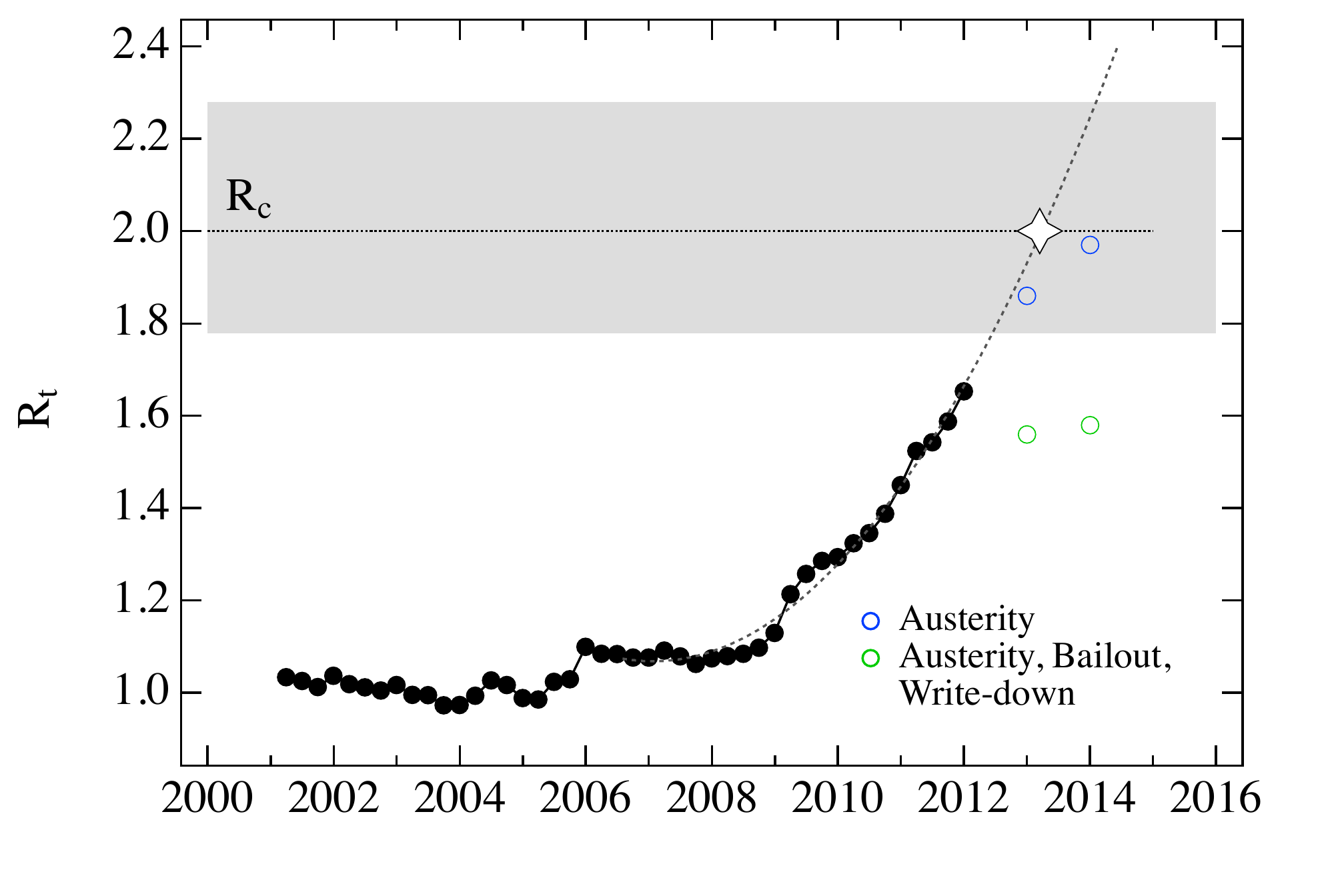}}
\caption{\textbf{Time of default: Greece} - We obtain the time of default for Greece by
projecting Greece's debt ratio (solid dots) as shown (dashed line). Default becomes certain when the debt to GDP ratio reaches the value $R_c=2.0$ in March 2013. Default is preempted by the impact of the negotiations between Greece and the European Union including a $53.5\%$ haircut, austerity and ESPF loans as shown in projections for 2012 and 2013 (green circles) according to estimates reported in Ref. \cite{Baumann2012}. The impact of deficit targets without the other interventions led to much higher projections that did not avoid default (blue circles) \cite{Financial2012}. The horizontal shading corresponds to the range of debt ratios for defaults of 20-80\% recovery (see Fig. \ref{fig:Pfits}).}
\end{figure}

These two parameters can also be used to evaluate if the market interest rates of a country are
consistently set according to economic fundamentals.
The probability of default $P_t$ increases as the interest rates $i_t$ increase, and it goes to 0 as they approach the risk-free interest rates $r_t$, according to \cite{Calvo1989}

\begin{equation}
P_t = \left(1-\frac{1+r_t}{1+i_t}\right)\frac{1}{1-\rho}
\label{eq:pdefault_text}
\end{equation}
Equating Eq. \ref{eq:final_p_text} and \ref{eq:pdefault_text} we have
\begin{equation}
\displaystyle
R_t = R_c-\eta X_t
\label{eq:linear_intext}
\end{equation}
where $X_t=\ln(1+r_t)-\ln(i_t-r_t)$ (see Appendix \hyperref[app:a]{A} for details). Therefore, if the experimental relationship between $R_t$ and $X_t$ is linear, international markets are
consistently setting interest rates according to the debt ratio (see Fig. \ref{fig:equilibrium}).


\section{results}
We tested the model on Greece over the last decade. Results are shown in Fig. \ref{fig:Pfits} assuming a recovery rate of 50\%, the average value on defaulted sovereign bonds over the last 30 years \cite{Moodys2010}. (The model parameters do not depend on the recovery rate used.) The figure shows the probability of default plotted against the debt ratio. As predicted by Eq. \ref{eq:final_p_text}, the data lies on a sigmoidal curve, and the two model parameters can be extracted from this fit. We obtain a critical debt ratio $R_c=2.0$, and a heterogeneity parameter $\eta=0.18$. Therefore, the model estimates that when Greece's debt reaches a level about twice its GDP, the market projects the country default probability to be 1.

Figure \ref{fig:poly_fit} shows how the time of certain default of a country can be estimated. We approximate Greece's debt trajectory in one of two ways, using a numerical polynomial fit or an economic debt projection (Appendix \hyperref[app:b]{B}). Either way, the debt ratio would reach the default threshold $R_c$ in the first 
few months of 2013. Simulations with different intervention scenarios (Appendix \hyperref[app:b]{D}) show that the default could not have been prevented solely by meeting the austerity targets recently agreed upon \cite{Financial2012}. Indeed, global fears of a Greek default led to negotiations and to a coordinated intervention that resulted in a haircut, which can be considered tantamount to the partial default projected by our model.
The debt write-down occurred at a time when the probability of default was 33\%, one year before the projected certain default. A separate analysis shows that the timing of default can be attributed to non-equilibrium market fluctuations \cite{Lagi2012}. 
Negotiated partial defaults are more likely than a full default because of the harm to both lenders and borrowers \cite{Fane1995}. In this case, negotiations led to a 53\% write-down of the debt by the private sector (in the form of a waiver of receivables) coupled with access to EFSF and IMF loans at preset interest rates and austerity measures. If the austerity measures are met, and its impacts on GDP are consistent with published expectations (Appendix \hyperref[app:c]{C}), this will allow the debt ratio to remain below the danger zone.

We can use our results to evaluate whether market interest rates are consistent with equilibrium assumptions over time. During the near-default period, bond rates change rapidly and are subject to higher uncertainty. In this scenario, the impact of large traders and trend following may become relevant. The fit of the default dynamics for Greece shown in Figures \ref{fig:Pfits} and \ref{fig:fits} is a test of market self-consistency: if the market did not set interest rates consistently with the valuation of the critical debt ratio in prior years, these relationships would not hold over time. For example, if the interest rates overestimated the probability of default of a country in a given year relative to the historical precedent, the corresponding point would lie above the fit in Fig. \ref{fig:Pfits}. The self-consistency condition can also be represented as a linear relationship between debt ratio and the distance of market interest rates from their default value (see Fig. \ref{fig:fits}). The good fits imply that the reference 10 year bond interest rates averaged over yearly time frames are consistent with market equilibrium. Shorter time frame behaviors deviate from these conclusions and are treated in a separate paper \cite{Lagi2012}.

We also tested the model for the other European countries involved in the debt crisis: Portugal, Ireland, Spain and Italy (see Fig. \ref{fig:portugal}-\ref{fig:italy} in Appendix \hyperref[app:d]{D}). As in the case of Greece, we found that markets self-consistently determined a definite value for the critical debt ratio $R_c$ for all countries. The time range over which a fit with a consistent value of $R_c$ is possible depends on the country: the model fits the last 11 years for Greece and Italy, 
9 years for Portugal, and 5 years for Spain and Ireland.

As for the case of Greece, the equilibrium model portrays a situation which could be less severe than the one described by secondary markets and media \cite{Oakley2012, Post2012, Kavoussi2011, Elliott2011}. If the budget targets are met, Portugal could avoid a partial default at the beginning of 2013. Ireland and Spain have a larger buffer before a default, projected for mid-2014. Italy is still
farther away from a possible default, projected by the model for mid-2016, if the current trend is maintained. In all cases an immediate large scale policy response to change the debt trajectory does not seem to be necessary.

The values of the critical debt ratio, $R_c$, do not vary over orders of magnitude between countries. The values cluster around unity, which could imply that lenders believe a country is able to repay its debt when the debt is about the size of its GDP, but not significantly larger. The capacity of a country to borrow depends on a variety of factors, and reflects creditors' concerns about solvency (long-run prospects), liquidity (short-run prospects) and repudiation risks \cite{Cooper1984}. The reasons why countries usually borrow lower-than-sustainable amounts, i.e. why $R_t$ is kept well below $R_c$, have been discussed \cite{Cooper1984}. Even though we find there is only one scale for this parameter, $R_c$ varies among countries. The value of $R_c$ may depend on a variety of macroeconomic phenomena. Our analysis does not identify these dependencies. Sustainable growth rate, the coupling to other economies willing to help, considerations about the country's hidden economy and its possibility to increase taxes are also factors that might contribute to the country-specific value of $R_c$.

Finally, it is worth noting that the framework we presented here is a long-run model of default for European countries, but short timescales require a different picture. 
This is demonstrated by the deviation of 10-year Greek bond prices from their equilibrium trajectory after the summer of 2011, 
when they dropped from 57\% to 21\% of their face value 
in five months. One might think that this increase in interest rates would be justified by the default projections, even if they were about 1.5 years away according to our equilibrium model. But the corresponding probability of default at that time was around 0.3 (see Fig. \ref{fig:Pfits}), not large enough to explain such a high level of interest rates, nor to justify an imminent write-down.
The equilibrium economics perspective is uncompromising: bond prices should follow the curve regardless of fluctuations---the trajectory has been defined. In contrast, a 
complex systems perspective predicts that once the system is near a critical point, multiple stable states (in this case default and non-default) become possible. In this regime, the market is not self-averaging, its finite size gives rise to fluctuations at all scales, and sensitivity to trend following and market manipulation increases. 
Both equilibrium and non-equilibrium mechanisms are therefore included in our framework and are not inconsistent, as they are acting at different timescales.


\section{conclusions}
We developed a quantitative model of sovereign default for European countries that identifies the debt ratio as the fundamental parameter that investors use to set the persistent value of interest rates captured by annually-averaged values.
Our analysis shows that the equilibrium model assumption that interest rates are set by economic conditions is satisfied,
though we do not show that interest rates actually reflect potential losses due to default risk. 
Our model is able to address several questions. First, it gives an estimate of the debt sustainability of a country, i.e. how much creditors are willing to lend. In the case of Greece the maximum debt is twice the GDP, $R_c \sim 2.0$. Second, given the debt trajectory of the country, the model can identify the time the debt ratio goes beyond the threshold, triggering a default. Since default is itself determined by the time of diverging market interest rates, default is self-consistently determined by the market evaluation of the default probability. Third, the model can establish whether bond markets satisfy the equilibrium assumption that interest rates the market sets have a consistent relationship with fundamental indicators over time. We find this to be the case for Greece over the last 10 years. 

The model identifies the time of certain default as occurring early in 2013. While this appears to be in reasonable agreement with the debt write-down, the timing is different by approximately a year and this is critical to a proper interpretation of events. The existence of non-equilibrium bandwagon effects that preempt the equilibrium default means that higher national interest rates and austerity measures 
than would be justified in equilibrium accelerated and exacerbated the default process. Without these effects, more modest measures and less social disruption could have been sufficient to avert default. 

As it stands today, drastic measures taken by the government, international organizations and the private sector may avoid a complete default. These interventions have both an economic and social cost. By regulating markets to limit the non-equilibrium market fluctuations, the cost of interventions and their severity could be reduced for Greece and other European countries.


\section{acknowledgements}
We thank Karla Z. Bertrand for help with the literature review, Jeffrey Fuhrer, Richard Cooper, Yavni Bar-Yam and Dominic Albino for helpful comments.


\newpage
\phantomsection
\label{app:a}
\begin{center}
\Large Appendix A\\
\Large Model Details
\end{center}

We consider a country that decides to borrow from international capital markets at interest rate $i$. At every time period $t$, the country is in one of two states: \emph{default} or \emph{nondefault}. Since the ability of the country to pay its debt $D$ depends on its economic output $Y$, the state of the country is determined by its debt ratio, $R = D/Y$. When $R$ is above a critical threshold $R_c$, the value of which depends on the cost of the penalties associated with defaulting, the country is in the default state, and vice versa.

If $P$ is the probability of default with recovery rate $\rho$, i.e. the probability that the country will pay a share of the debt $\rho < 1$, the expected return from a one-period loan would be $(1+i)(1-P+P \rho)$ \cite{Calvo1989}. By definition, $(1+i)$ is the contractual repayment at the end of the period per unit of loan and $(1-P)$ is the probability of full repayment. An investor, however, has the option of investing in risk-free bond markets and receive an expected return of $(1+r)$ at the end of the period, where $r$ is the risk-free interest rate. Thus, since in competitive equilibrium investors should be indifferent between those two alternatives, equating the two expressions for expected return and solving for $P$ we have

\begin{equation}
P = \left(1-\frac{1+r}{1+i}\right)\frac{1}{1-\rho}
\label{eq:pdefault}
\end{equation}
This equation establishes a relationship between the expected probability of default and interest rates. It is consistent with the intuition that the probability of default $P$ increases as the interest rates $i$ increase, and it goes to 0 as they approach the risk-free interest rates. Equation \ref{eq:pdefault} can also be written as a function of bond prices, $B$. For a one-period loan,

\begin{equation}
B = \frac{B_0(1+i_0)}{1+i}
\label{eq:bond}
\end{equation}
where $B_0$ is the initial value of the bond (face value) and $i_0$ the initial interest rate. Then, substituting in Eq. \ref{eq:pdefault} we get

\begin{equation}
P = \left(1-\frac{B}{B_0}\frac{1+r}{1+i_0}\right)\frac{1}{1-\rho}
\label{eq:pdefault_bond}
\end{equation}
which relates the probability of default with the current bond price.

There are three reasons why $i$ is in general greater than $r$: transaction costs, risk aversion and default risk \cite{Calvo1989}. For simplicity, let's assume that transaction costs are small enough to be neglected and that lenders are risk neutral. If lenders had perfect knowledge of the time of default of the country, and the 
duration of the one-period loan is short enough to be neglected,
they could keep $i=r$ until the debt reaches the critical default value $R_c$.  
At the time of default, when the debt reached the critical default value, 
the interest rates would become infinite. Therefore, at each time $t$ this idealized bistable system can be summarized as a step function of the default probability:

\begin{equation}
\left \{ \begin{array}{l}
P_t = 0, \hspace{2ex} i_t = r_t \hspace{7ex} R_t < R_c \\ \\
P_t = 1, \hspace{2ex} i_t = \infty  \hspace{6ex} R_t > R_c
\end{array}
\right.
\label{eq:bistable}
\end{equation}
\vspace{1ex}

\begin{figure}[t]
\refstepcounter{figref}\label{fig:sigmoid}
\href{http://www.necsi.edu/research/economics/bondprices/sigmoid.pdf}{\includegraphics[width=0.9\linewidth]{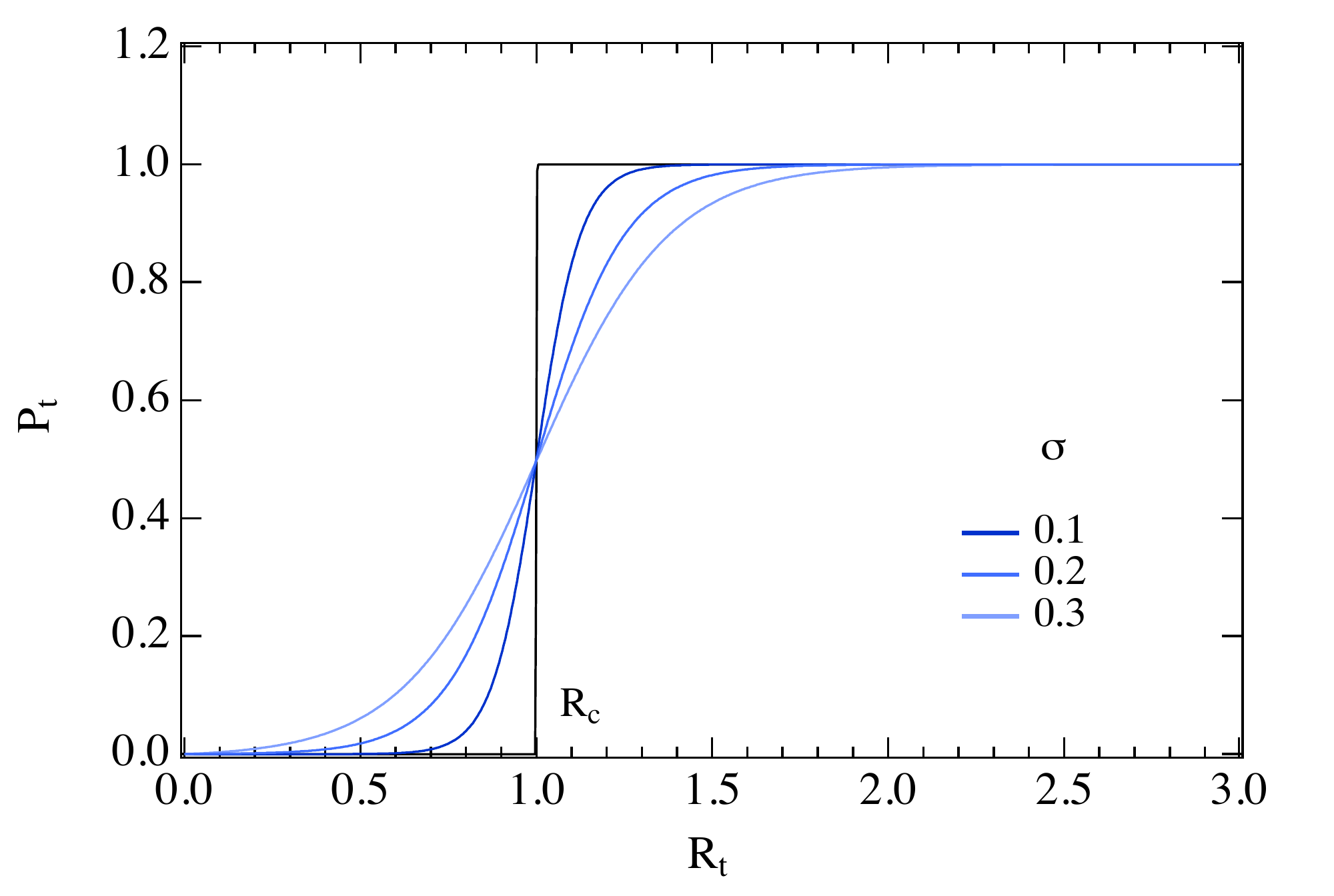}}
\caption{\textbf{Default probability} - Dynamics of the probability of sovereign default as a function of the country's debt ratio, assuming no recovery. In an idealized scenario in which lenders have perfect information, a step function of the default probability would occur (black line, Eq. \ref{eq:bistable}). Imperfect knowledge and heterogeneous strategies results in smoothed sigmoidal function (blue lines, Eq. \ref{eq:final_p}). The extent of smoothing depends on the width of the distribution of expectations given by $\sigma$.}
\end{figure}

We plot the default probability for the case $R_c=1$ in Fig. \ref{fig:sigmoid} (black curve). This step function is in practice smooth (blue curves), since the investment strategies are heterogeneous for a number of reasons: limited amount of information available to market participants, different estimates of the critical debt ratio, imperfect knowledge of the debt trajectory, different evaluations of the political decision-making process behind a possible default, different bond maturation periods and payment schedules, disparate influence of other economic indicators. Here we show how this smooth behavior naturally arises from a supply and demand model of capital markets that includes heterogeneous expectations. The result follows from the reasonable assumption that all dependencies are smooth other than the singular behavior of the default itself. Smoothness in the vicinity of the default justifies expanding all functional relationships to first order. The following derivation is illustrative and captures the eventual behavior.

If we assume a linear relationship between bond prices $B$ and quantity of debt supplied by the country ($Q_s$) or demanded by investors ($Q_d$) we have

\begin{equation}
\displaystyle
Q_s = \alpha_s + \beta_sB
\label{eq:qs}
\end{equation}
\begin{equation}
\displaystyle
Q_d = \alpha_d - \beta_dB
\label{eq:qd}
\end{equation}
and, since at equilibrium $Q_s = Q_d$,

\begin{equation}
\displaystyle
B_t = \frac{\alpha_d(t)-\alpha_s(t)}{\beta_d+\beta_s}
\label{eq:ieq}
\end{equation}
assuming that supply and demand shocks would influence the intercepts---and not the slopes---of the linear dependences. The demand intercept $\alpha_d$ depends on the overall liquidity available for financing the country's debt, and therefore on the number of market participants. Rational investors are willing to lend money to the country as long as its debt ratio is below their estimate of $R_c$. Due to all of the factors contributing to heterogeneity, we expect investor estimates of the critical value of default to have a distribution, whose form
can be approximated by a normal distribution with average $R_c$ and standard deviation $\sigma$,

\begin{equation}
\displaystyle
f(R_t) = \frac{N}{\sigma\sqrt{2\pi}}\exp\left[-\frac{1}{2}\left(\frac{R_t-R_c}{\sigma}\right)^2\right]
\label{eq:gauss}
\end{equation}
where $N$ is the total number of investors. Since agents that estimate the default at a particular value of $R_t$ expect it also for any greater value of the debt ratio, $\alpha_d$ is proportional to the cumulative distribution function 

\begin{equation}
\displaystyle
\alpha_d (t) \sim  \frac{N}{2}\left[1-\erf\left(\frac{R_t-R_c}{\sigma\sqrt{2}}\right)\right]
\label{eq:alphas}
\end{equation}
where $\erf()$ is the error function. This expression can be well approximated with a sigmoid function \cite{Bryc2002}, according to 

\begin{equation}
\displaystyle
\alpha_d (t) \approx \frac{N}{1+e^{(R_t-R_c)/\eta}}
\label{eq:alphas_sig}
\end{equation}
where $\eta = \sigma\sqrt{2\pi}/4$ is the parameter characterizing the heterogeneity. The bond supply intercept, $\alpha_s$, depends on the amount of money a country wants to borrow, and thus on its debt ratio. We expand this dependence around the critical value $R_c$, keeping only the constant term. The effect of the absolute level of the debt ratio ($R_t$) on the supply (and more generally on bond prices) can be neglected compared to the effect of the sigmoidal function dependence ($R_t-R_c$), i.e the distance of the debt ratio from its critical value. Substituting Eq. \ref{eq:alphas_sig} into Eq. \ref{eq:ieq} we have

\begin{equation}
\displaystyle
B_t = k' \frac{1}{1+e^{(R_t-R_c)/\eta}} + k''
\label{eq:almost_final_b}
\end{equation}
where $k'$ and $k''$ are constants that are functions of $\alpha_s, \beta_s, \beta_d$ and $N$. Their value can be inferred by imposing constraints on two limiting cases of Eq. \ref{eq:bond}:

\begin{itemize}
\item $i_t \rightarrow \infty$, $B_t \rightarrow 0$ when $R_t \rightarrow \infty$
\item $i_t \rightarrow r_t$, $B_t \rightarrow B_0(1+i_0)/(1+r_t)$ when $R_t \rightarrow 0$
\end{itemize}
Using these conditions we obtain

\begin{equation}
\displaystyle
B_t = B_0\left(\frac{1+i_0}{1+r_t}\right)\frac{1+e^{-R_c/\eta}}{1+e^{(R_t-R_c)/\eta}}\sim B_0\left(\frac{1+i_0}{1+r_t}\right)\frac{1}{1+e^{(R_t-R_c)/\eta}}
\label{eq:final_b}
\end{equation}
since $e^{-R_c/\eta} \ll 1$. Inserting this expression in Eq. \ref{eq:pdefault_bond},

\begin{equation}
\displaystyle
P_t = \left[1-\frac{1}{1+e^{(R_t-R_c)/\eta}}\right]\frac{1}{1-\rho}
\label{eq:final_p}
\end{equation}
This equation relates the probability of default to the two parameters of our model, $R_c$ and $\eta$. The family of curves $P_t$ vs $R_t$ at $\rho=0$ for different values of $\sigma$ are shown in Fig. \ref{fig:sigmoid}. The asymptotic behavior of the probability of default approaching 1 when the debt ratio goes to infinity is justified by theoretical \cite{Eaton1981}, empirical \cite{Reinhart2003} and numerical data \cite{Yue2010}.

Equating Eq. \ref{eq:pdefault} with Eq. \ref{eq:final_p} gives

\begin{figure}[tb]
\refstepcounter{figref}\label{fig:fits}
\href{http://www.necsi.edu/research/economics/bondprices/line_fit.pdf}{\includegraphics[width=0.9\linewidth]{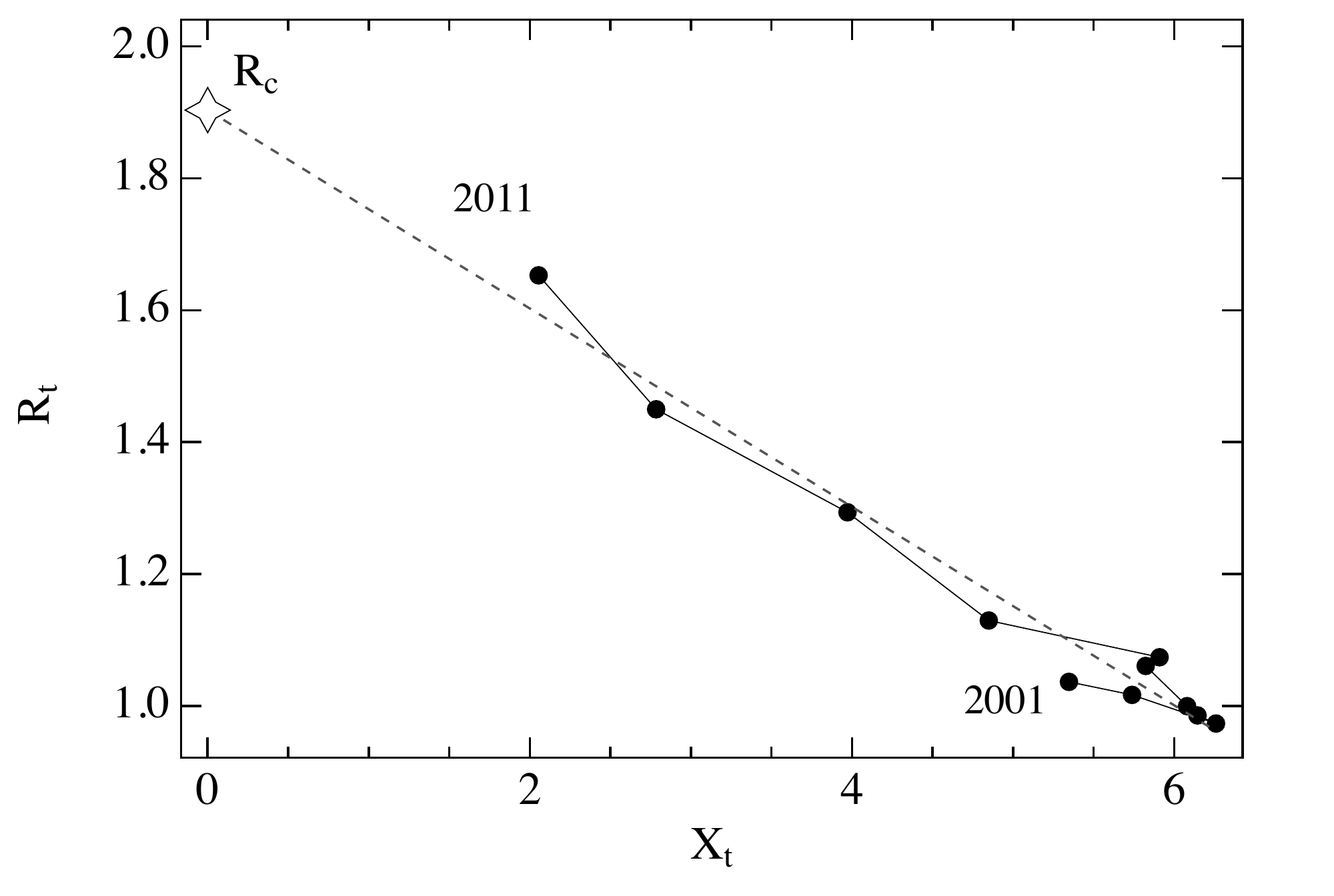}}
\caption{\textbf{Test of equilibrium market model of default} -  Greece's debt ratio \cite{eurostat} as a function of the default distance parameter $X_t$ (Eq. \ref{eq:linear2}) from 2001 to 2011 inclusive. $X_t$ reflects long-term interest rates, i.e. secondary market annually-averaged yields of 10-year bonds \cite{oecd}. Data (solid line) can be fitted with Eq. \ref{eq:linear2} (dashed line) to obtain the two model parameters, $R_c=1.90\pm0.05$ and $\eta = 0.15\pm0.01$. The small difference between these values and the ones obtained from the fitting in Fig. \ref{fig:Pfits} is due to the logarithmic scale. Data shown for the debt ratio are end-of-year estimates, while long-term interest rates are averaged over the course of the year. Using annually-averaged debt ratios or quarterly values for both debt ratio and interest rates does not significantly affect the results. Fitted parameters in all cases are the same within statistical uncertainty.}
\end{figure}

\begin{equation}
\displaystyle
R_t = R_c-\eta\ln\left(\frac{1+r_t}{i_t-r_t}\right)
\label{eq:linear1}
\end{equation}
When the debt ratio reaches the critical threshold, $R_t=R_c$, the corresponding critical value for interest rates is
\begin{equation}
\displaystyle
i_c = 1 + 2r_t \sim 1
\label{eq:ic}
\end{equation}
since risk-free interest rates are negligible compared to the interest rates of a country close to default. We can rewrite Eq. \ref{eq:linear1} as,

\begin{equation}
\displaystyle
R_t = R_c-\eta X_t
\label{eq:linear2}
\end{equation}
where $X_t$ is a measure of the distance between the current value of interest rates and the value at default. This linear relationship between $R_t$ and $X_t$ is verified empirically in Fig. \ref{fig:fits} for Greece, and in Appendix \hyperref[app:d]{D} for Portugal, Ireland, Spain and Italy. While the slope of the linear fit gives the heterogeneity parameter, its intercept is the critical debt ratio of the country.

The debt threshold $R_c$ does not depend on the assumed amount of debt recovery, but the time of expected default does. Eq. \ref{eq:final_p} specifies the probability of default for different debt thresholds and implicitly the debt ratio at which the default probability reaches one, $R_d=R_c+\eta \ln(1/\rho-1)$. For the generally observed case of 50\% recovery, the critical debt ratio coincides with the debt threshold, $R_d=R_c$.


\newpage
\phantomsection
\label{app:b}
\begin{center}
\Large Appendix B\\
\Large Debt Feedback Loop
\end{center}

In this Appendix, we provide a quantitative description of the feedback loop in Fig. \ref{fig:feedback} (blue arrows). We are going to use this analysis to identify the expected change of the debt ratio over time, given policy options that provide below market interest rates such as austerity and bailouts.

The country's debt trajectory is

\begin{equation}
D_t = S_t + D_{t-1}(i_t+1)
\end{equation}
where $D$ is the gross debt, $i$ is the market interest rates and $S$ is the primary deficit, i.e. government spending minus tax revenue. If we divide the previous equation by the economic output $Y_t$ we have

\begin{equation}
R_t = s_t + \frac{D_{t-1}}{Y_t}(i_t+1)
\end{equation}
where $s_t=S_t/Y_t$ is the budget ratio. If we define the GDP growth as $y_t = (Y_t - Y_{t-1})/Y_{t-1}$ and note that $D_{t-1}/Y_t = R_{t-1}/(1+y_t)$, from the previous equation we have the accumulation equation,

\begin{equation}
\displaystyle
R_t = s_t + R_{t-1}\left(\frac{i_t-y_t}{1+y_t}+1\right).
\label{eq:ratio}
\end{equation}
This equation completes the feedback loop represented in Fig. \ref{fig:feedback}: fundamental economic indicators determine the expected default probability (arrow 1 and Eq. \ref{eq:final_p}), which influences market interest rates (arrow 2 and Eq. \ref{eq:pdefault_bond}), which in turn influences the new value of economic indicators (arrow 3 and Eq. \ref{eq:ratio}). This feedback loop reflects the role of interest payments themselves on the increase in country debt.

\begin{figure}[tb]
\refstepcounter{figref}\label{fig:accumulation}
\href{http://www.necsi.edu/research/economics/bondprices/accumulation_fit.pdf}{\includegraphics[width=0.9\linewidth]{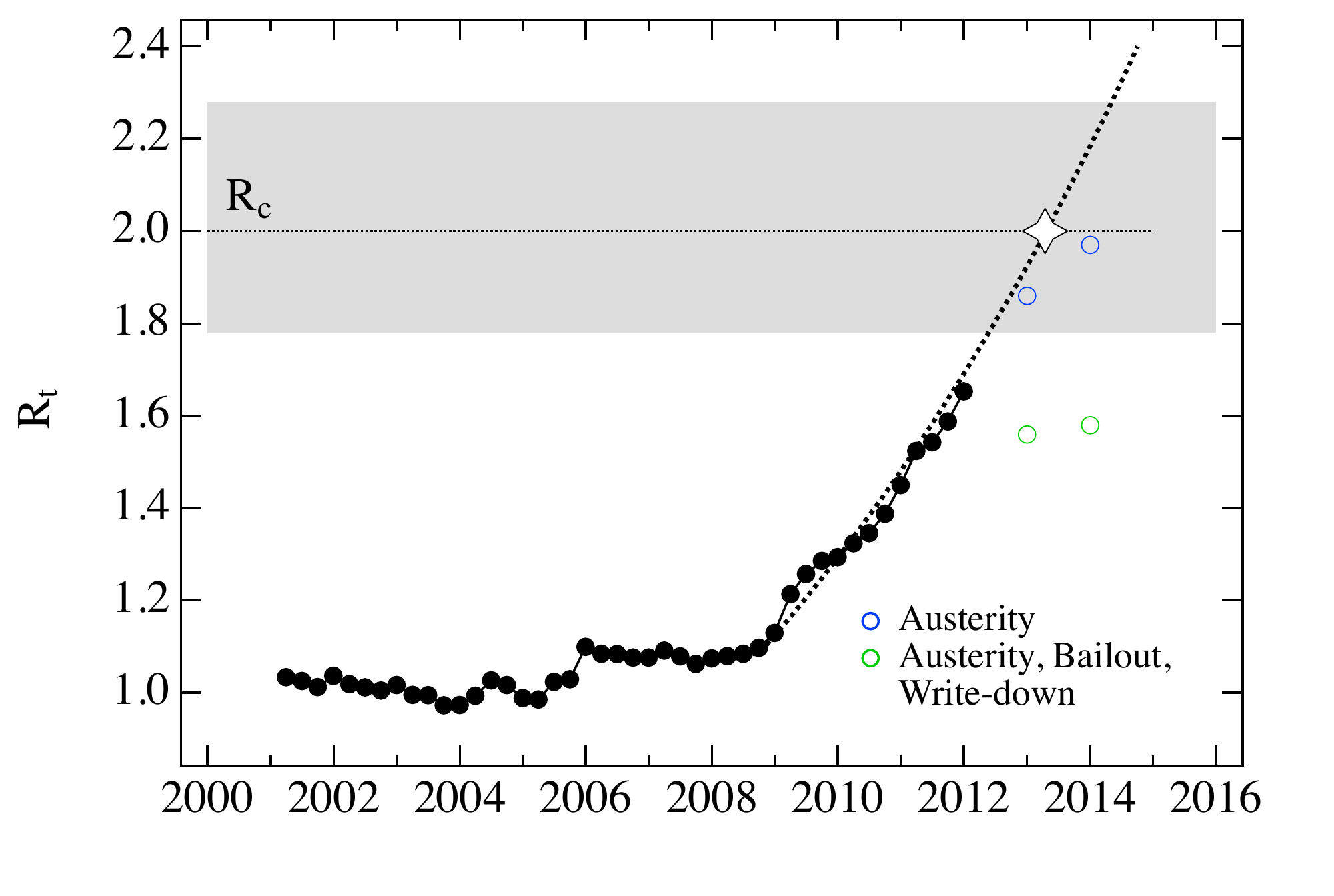}}
\caption{\textbf{Extrapolation} - Greece debt ratio trajectory projection using Eq. \ref{eq:differential_sol} and assuming time-independent quantities, with quarterly values: $i = 0.021$, $s = 0.008$ and $y = -0.007$.}
\end{figure}

Finally, we show how we can solve Eq. \ref{eq:ratio} and fit the debt trajectory analytically. The following equations, which are not used anywhere else in the paper or in the default model, demonstrate that in order to avoid default in the 
not-too-long term, the GDP has to grow faster than borrowing. As we saw in Eq. \ref{eq:ratio}, interest rates can affect the debt ratio, and the debt ratio grows very rapidly if $i_t > y_t$, i.e. if the interest rates are larger than the rate of economic growth. To obtain a closed-form solution that shows the dependence of the debt ratio on the GDP change and interest rates, we assume that $i_t$, $y_t$ and $s_t$ are not time dependent---this assumption is only used for the analytic fit in this Appendix, and not for our model or elsewhere in this paper---and solve Eq. \ref{eq:ratio} to give

\begin{equation}
\displaystyle
R_t = R_0\left(\frac{1+i}{1+y}\right)^t + s\frac{1+y}{i-y}\left[\left(\frac{1+i}{1+y}\right)^t-1\right] 
\label{eq:difference_sol}
\end{equation}
where $R_0$ is the value of the debt ratio at time 0. The difference equation \ref{eq:ratio} can be transformed into a differential equation if the usual assumption of smoothness applies,

\begin{equation}
\displaystyle
\frac{dR(t)}{dt} + \frac{y-i}{1+y}R(t) = s.
\end{equation}
This equation can be solved to give

\begin{equation}
\displaystyle
R(t) = R_0\exp\left(\frac{i-y}{1+y}t\right) + s\frac{1+y}{i-y}\left[\exp\left(\frac{i-y}{1+y}t\right)-1\right] 
\label{eq:differential_sol}
\end{equation}
Equation \ref{eq:difference_sol} coincides with Eq. \ref{eq:differential_sol} for small values of $i-y$, and either demonstrates how the debt ratio grows exponentially if $i > y$. Since our model shows that there is a threshold debt ratio set by markets, an exponential growth is unsustainable. The fit of the time dependence of the Greek debt ratio with Eq. \ref{eq:differential_sol} is shown in Fig. \ref{fig:accumulation}.
We emphasize that Equation \ref{eq:ratio}, which includes the time dependence of $i_t$, $y_t$ and $s_t$, is more general than Equation \ref{eq:differential_sol}. The latter is only used in this Appendix and not elsewhere in this paper.


\newpage
\phantomsection
\label{app:c}
\begin{center}
\Large Appendix C\\
\Large Austerity Measures
\end{center}

In order to evaluate the effect of austerity measures on the debt ratio of a country, we use Eq. \ref{eq:ratio} to calculate $R_t$ given estimates by economic analysts for the GDP change $y_t$ and austerity targets for the budget ratio $s_t$. 
We assume the interest rates for the next two years are similar to current rates, $i_t\sim i_{t-1}$, which is justified by the small absolute difference over time of the short-term interest rates, and the very few years over which the projection is being calculated. 
Our purpose is solely to evaluate whether the magnitude of austerity measures that have been implemented may be able to avert a default, using available economic projections according to the debt threshold we obtained.
We report the estimates for $y_t$ and the targets for $s_t$ in Table \ref{tab:values}.
The impact of the austerity programs on debt ratios are incorporated into the figures in Appendix D, showing their implications for averting defaults. \\

\begin{table} [h]
\center
	\begin{tabular}{| c | c | c |c | c | c|}	
	\hline
\emph{\textbf{Country}}	 &	\emph{\textbf{$s_{2012}$}}	 &	\emph{\textbf{$s_{2013}$}}	 &	\emph{\textbf{$y_{2012}$}}	&	\emph{\textbf{$y_{2013}$}} &	\emph{bailout}\\ 
	\hline
Greece	& 7.3	 \cite{Financial2012}	& 4.7 \cite{Financial2012} &  -6.4 \cite{Baumann2012} & -1.9 \cite{Baumann2012} & yes\\
	\hline
Portugal	& 4.5	 \cite{kowsmann2011}	& 3.0 \cite{kowsmann2011} &  -3.4 \cite{BancoDePortugal2012} & 0.0 \cite{BancoDePortugal2012} & yes\\
	\hline
Ireland	& 8.6	 \cite{RTE2012}	& 3.0 \cite{OCarroll2010} &  0.7 \cite{Finfacts2012} & 2.2 \cite{Thejournal2012} & yes\\
	\hline
Spain	& 5.3	 \cite{Reis2012}	& 3.0 \cite{Reis2012} &  -3.0 \cite{RossThomas2012} & -1.9 \cite{RossThomas2012} & no\\
	\hline
Italy		& 1.7	 \cite{irishtimes2012}	& 0.5 \cite{Donadio2012} &  -1.2 \cite{Brunsden2012} & 0.5 \cite{Donadio2012} & no\\
	\hline
	\end{tabular}
\caption{\textbf{Values used for debt ratio projections} - Austerity targets for the budget ratio $s_t$ and projections of the GDP change $y_t$ for the five countries 
we considered.}
\label{tab:values}
\end{table}


\newpage
\phantomsection
\label{app:d}
\begin{center}
\Large Appendix D\\
\Large Other Countries
\end{center}

\subsection{Portugal}
\begin{figure}[h]
\refstepcounter{figref}\label{fig:portugal}
\href{http://www.necsi.edu/research/economics/bondprices/portugal1.pdf}{\includegraphics[width=0.49\linewidth]{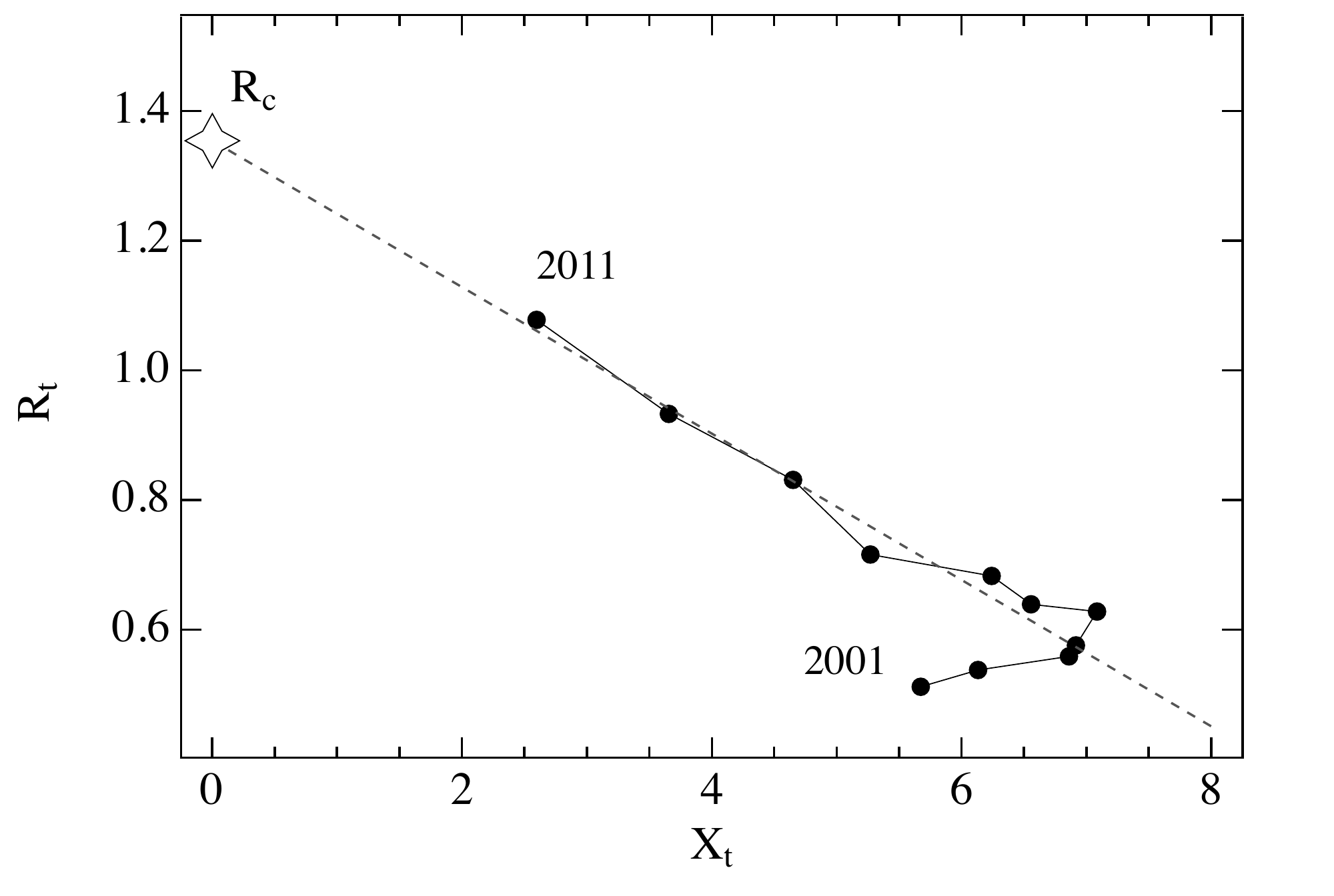}}
\href{http://www.necsi.edu/research/economics/bondprices/portugal2.pdf}{\includegraphics[width=0.49\linewidth]{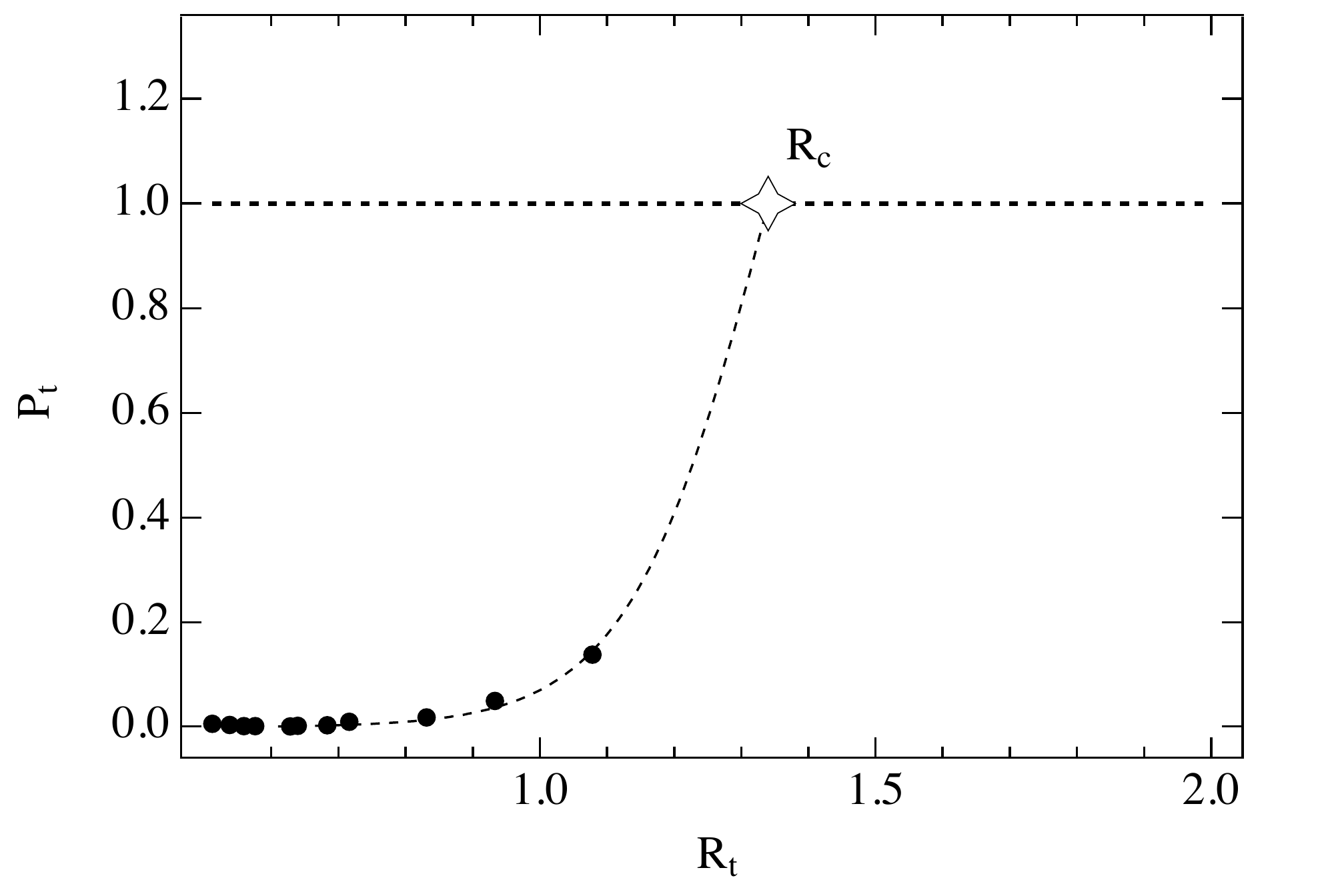}}
\href{http://www.necsi.edu/research/economics/bondprices/portugal3.pdf}{\includegraphics[width=0.49\linewidth]{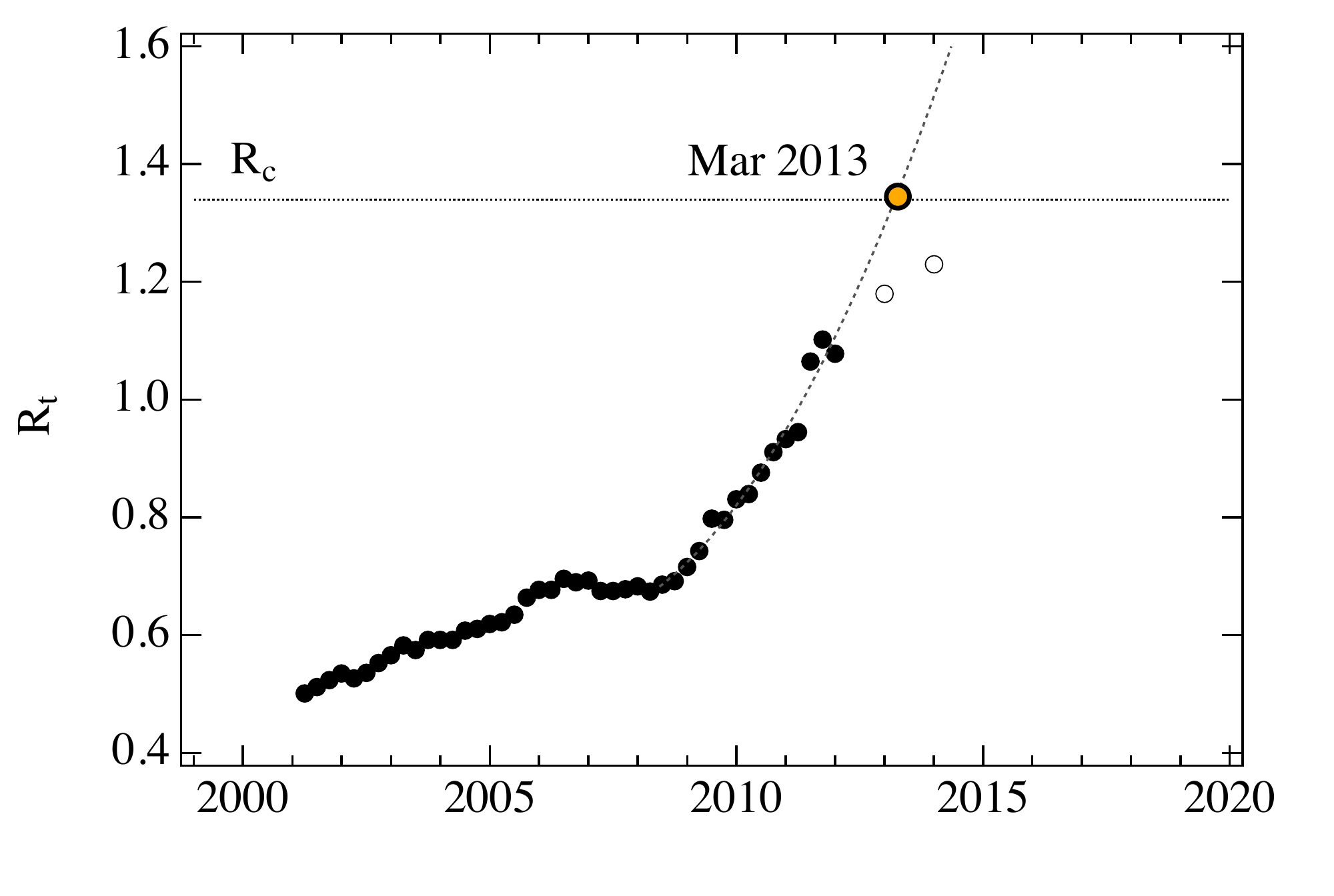}}
\caption{\textbf{Model fits and projections for Portugal} - \emph{Top left panel}: Portugal's debt ratio \cite{eurostat} as a function of the default distance parameter $X_t$ (Eq. \ref{eq:linear2}) from 2001 to 2011 inclusive. $X_t$ reflects long-term interest rates, i.e. secondary market yields of 10-year bonds \cite{oecd}. Data (solid line) can be fitted with Eq. \ref{eq:linear2} (dashed line) from 2007 to 2011 to obtain the two model parameters, $R_c=1.35\pm0.05$ and $\eta = 0.11\pm0.01$. \emph{Top right panel}: Probability of default as a function of Portugal's debt ratio. $P_t$ is calculated from Eq. \ref{eq:pdefault}, setting the recovery rate to $\rho =0.5$ and considering German interest rates as the risk-free interest rates. Data (dots) can be fitted with Eq. \ref{eq:final_p} (dashed line). This allows us to extract the two model parameters, the critical debt ratio $R_c=1.34\pm0.07$ and the heterogeneity parameter $\eta = 0.10\pm0.02$. \emph{Bottom panel}: Portugal's debt ratio (solid dots) as a function of time. The debt trajectory, projected with a third-order polynomial regression (dashed line), intersects the critical debt ratio ($R_c=1.34$) in Mar 2013. Hollow circles are projected debt ratios for the end of 2012 and 2013 if austerity measures are met (calculated from \cite{kowsmann2011,BancoDePortugal2012}).}
\end{figure}

\newpage
\subsection{Ireland}
\begin{figure}[h]
\refstepcounter{figref}\label{fig:ireland}
\href{http://www.necsi.edu/research/economics/bondprices/ireland1.pdf}{\includegraphics[width= 0.49\linewidth]{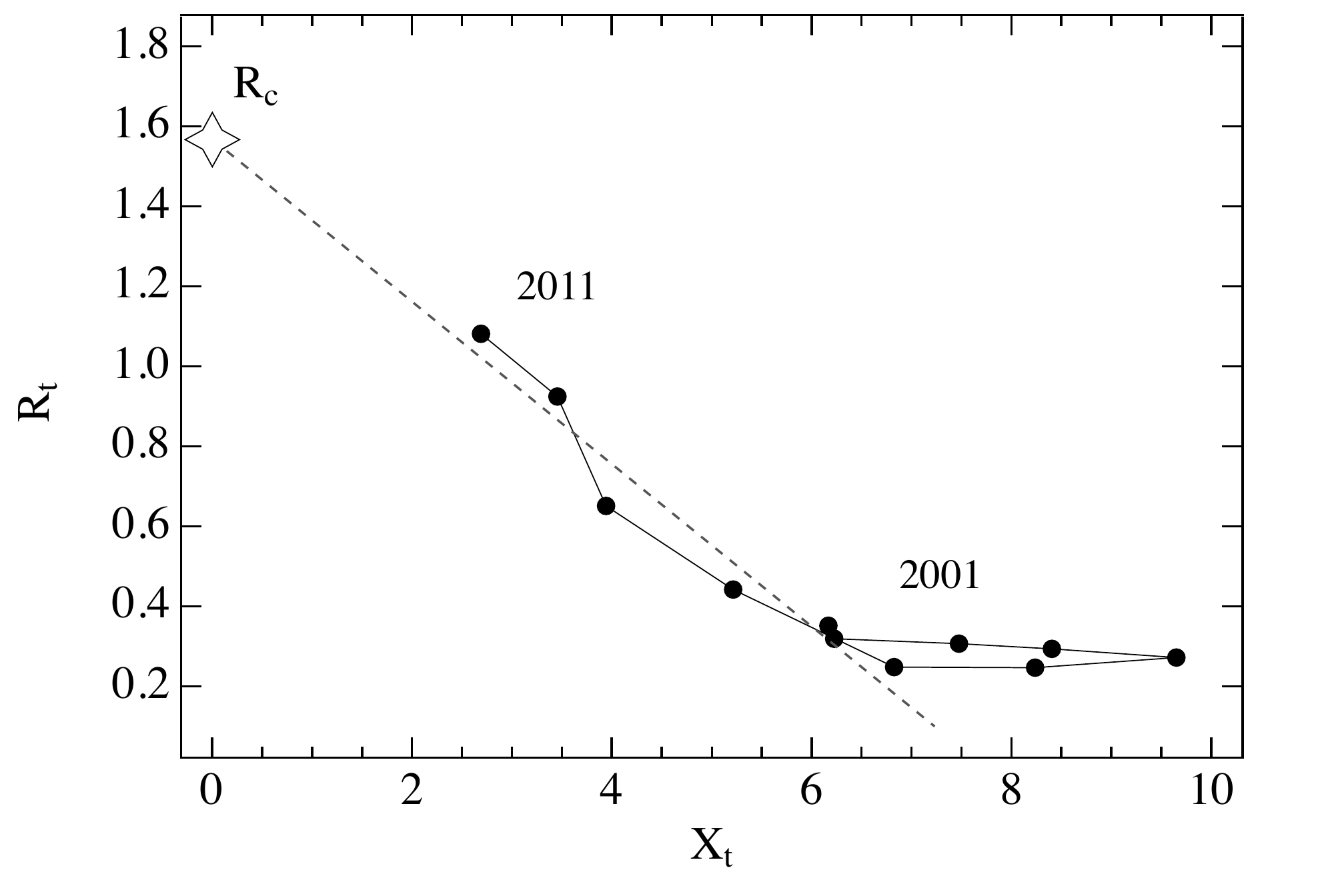}}
\href{http://www.necsi.edu/research/economics/bondprices/ireland2.pdf}{\includegraphics[width= 0.49\linewidth]{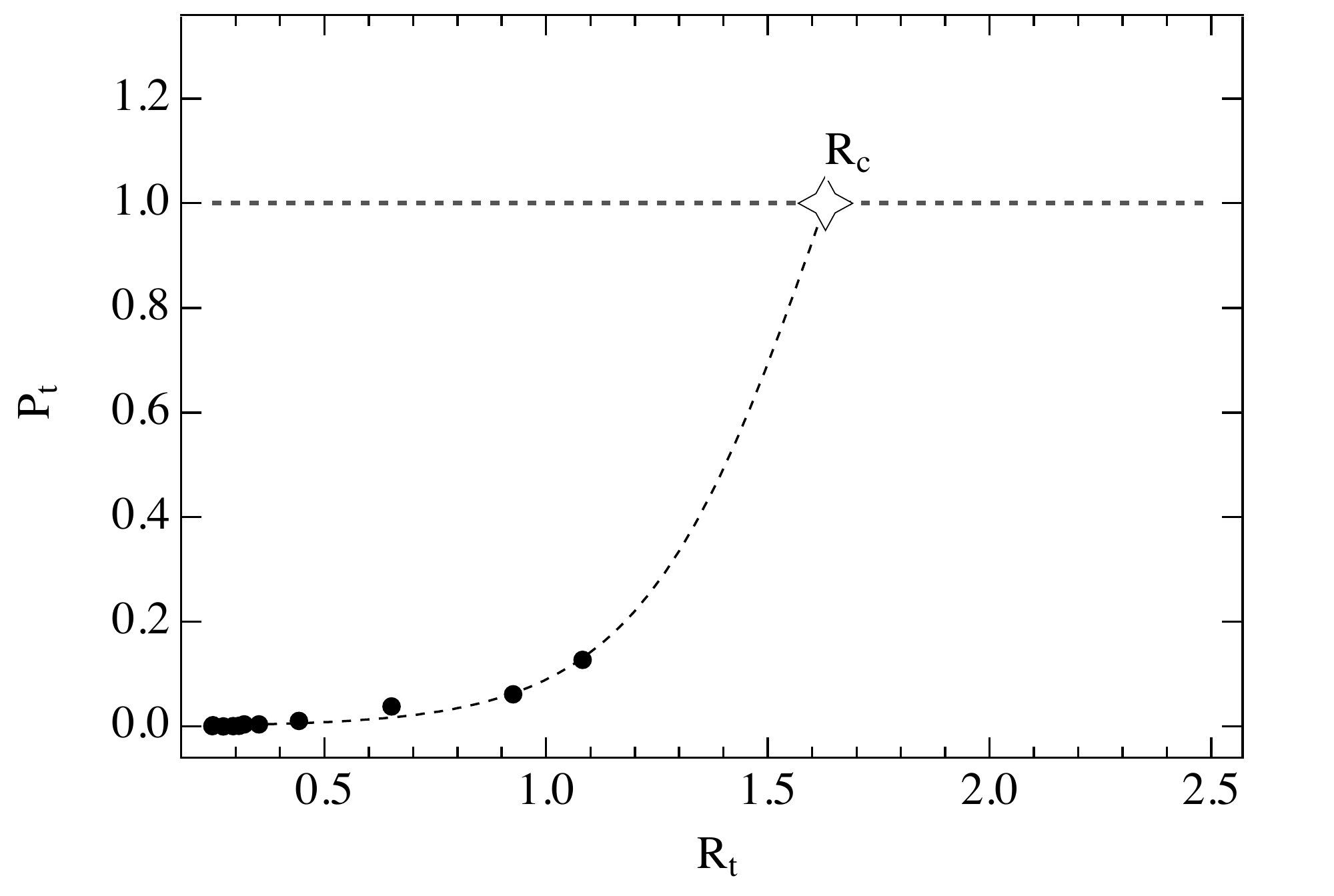}}
\href{http://www.necsi.edu/research/economics/bondprices/ireland3.pdf}{\includegraphics[width= 0.49\linewidth]{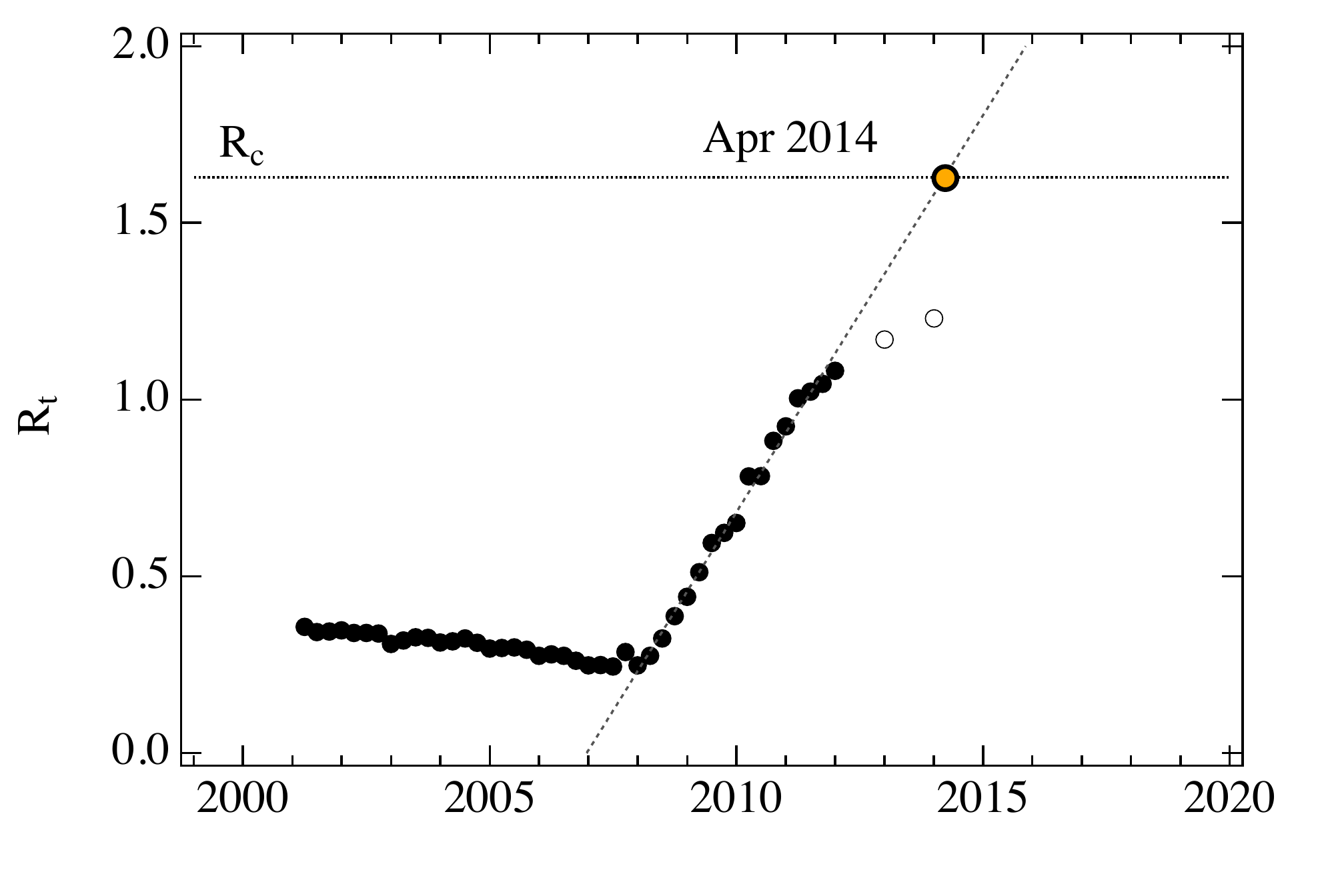}}
\caption{\textbf{Model fits and projections for Ireland} - \emph{Top left panel}: Ireland's debt ratio \cite{eurostat} as a function of the default distance parameter $X_t$ (Eq. \ref{eq:linear2}) from 2001 to 2011 inclusive. $X_t$ reflects long-term interest rates, i.e. secondary market yields of 10-year bonds \cite{oecd}. Data (solid line) can be fitted with Eq. \ref{eq:linear2} (dashed line) from 2007 to 2011 to obtain the two model parameters, $R_c=1.57\pm0.14$ and $\eta = 0.20\pm0.03$. \emph{Top right panel}: Probability of default as a function of Ireland's debt ratio. $P_t$ is calculated from Eq. \ref{eq:pdefault}, setting the recovery rate to $\rho =0.5$ and considering German interest rates as the risk-free interest rates. Data (dots) can be fitted with Eq. \ref{eq:final_p} (dashed line). This allows us to extract the two model parameters, the critical debt ratio $R_c=1.63\pm0.10$ and the heterogeneity parameter $\eta = 0.21\pm0.03$. \emph{Bottom panel}: Ireland's debt ratio (solid dots) as a function of time. The debt trajectory, projected with a linear regression (dashed line), intersects the critical debt ratio ($R_c=1.63$) in Apr 2014. Hollow circles are projected debt ratios for the end of 2012 and 2013 if austerity measures are met (calculated from \cite{RTE2012,OCarroll2010,Finfacts2012,Thejournal2012}).}
\end{figure}

\newpage
\subsection{Spain}
\begin{figure}[h]
\refstepcounter{figref}\label{fig:spain}
\href{http://www.necsi.edu/research/economics/bondprices/spain1.pdf}{\includegraphics[width=0.49\linewidth]{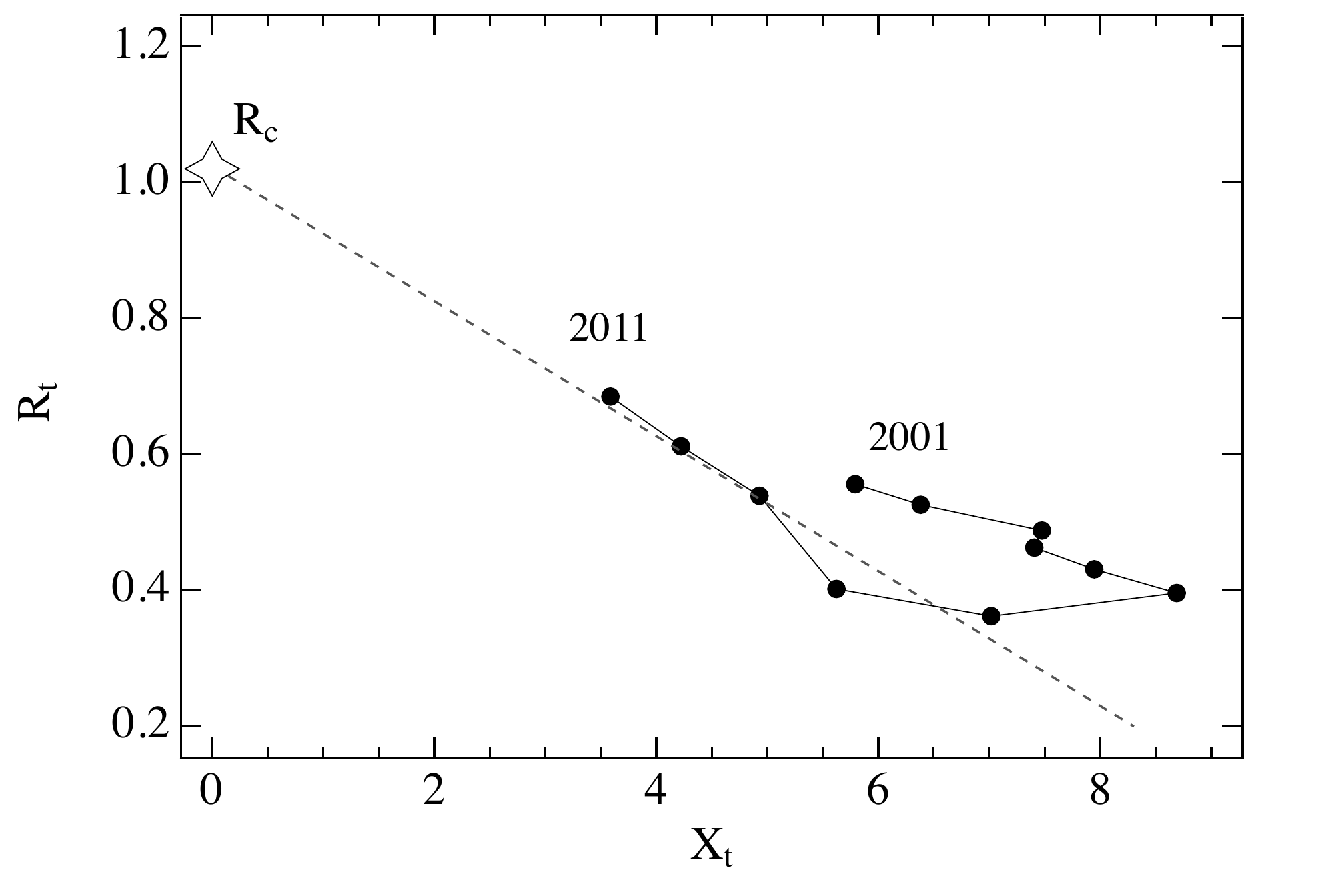}}
\href{http://www.necsi.edu/research/economics/bondprices/spain2.pdf}{\includegraphics[width=0.49\linewidth]{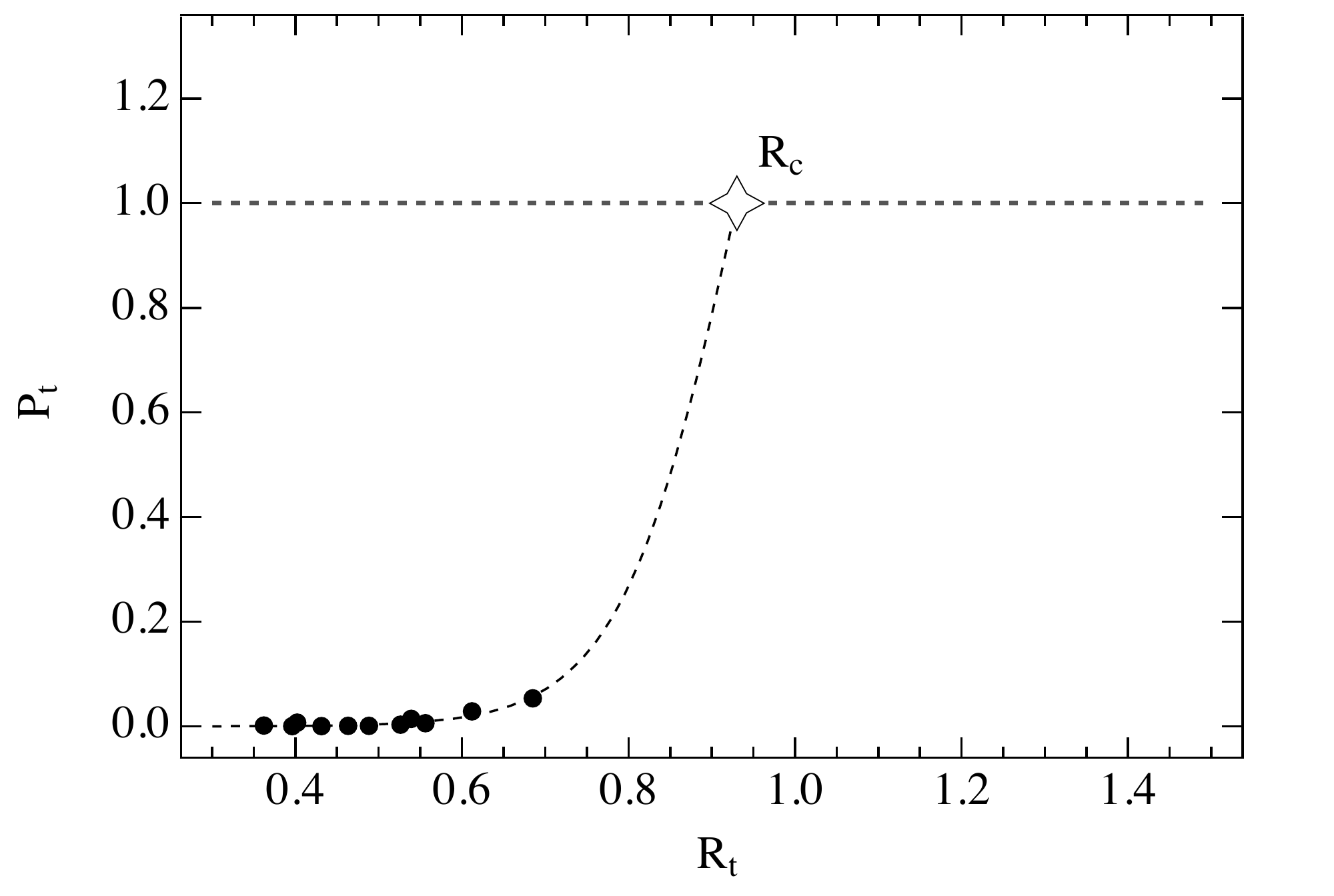}}
\href{http://www.necsi.edu/research/economics/bondprices/spain3.pdf}{\includegraphics[width=0.49\linewidth]{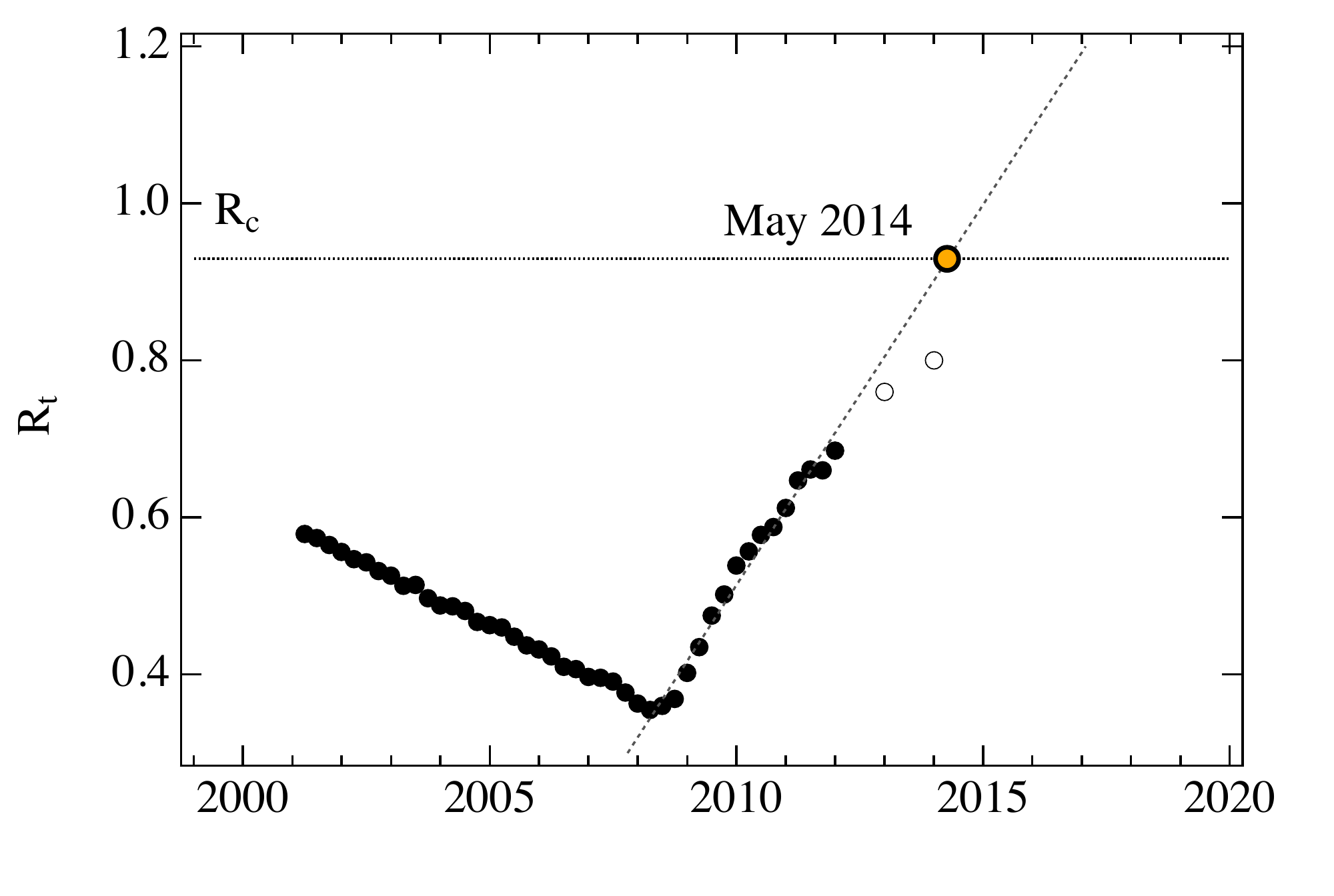}}
\caption{\textbf{Model fits and projections for Spain} - \emph{Top left panel}: Spain's debt ratio \cite{eurostat} as a function of the default distance parameter $X_t$ (Eq. \ref{eq:linear2}) from 2001 to 2011 inclusive. $X_t$ reflects long-term interest rates, i.e. secondary market yields of 10-year bonds \cite{oecd}. Data (solid line) can be fitted with Eq. \ref{eq:linear2} (dashed line) from 2007 to 2011 to obtain the two model parameters, $R_c=1.02\pm0.08$ and $\eta = 0.10\pm0.01$. \emph{Top right panel}: Probability of default as a function of Spain's debt ratio. $P_t$ is calculated from Eq. \ref{eq:pdefault}, setting the recovery rate to $\rho =0.5$ and considering German interest rates as the risk-free interest rates. Data (dots) can be fitted with Eq. \ref{eq:final_p} (dashed line). This allows us to extract the two model parameters, the critical debt ratio $R_c=0.93\pm0.05$ and the heterogeneity parameter $\eta = 0.07\pm0.01$. \emph{Bottom panel}: Spain's debt ratio (solid dots) as a function of time. The debt trajectory, projected with a linear regression (dashed line), intersects the critical debt ratio ($R_c=0.93$) in May 2014. Hollow circles are projected debt ratios for the end of 2012 and 2013 if austerity measures are met (calculated from \cite{Reis2012,RossThomas2012}).}
\end{figure}

\newpage
\subsection{Italy}
\begin{figure}[h]
\refstepcounter{figref}\label{fig:italy}
\href{http://www.necsi.edu/research/economics/bondprices/italy1.pdf}{\includegraphics[width=0.49\linewidth]{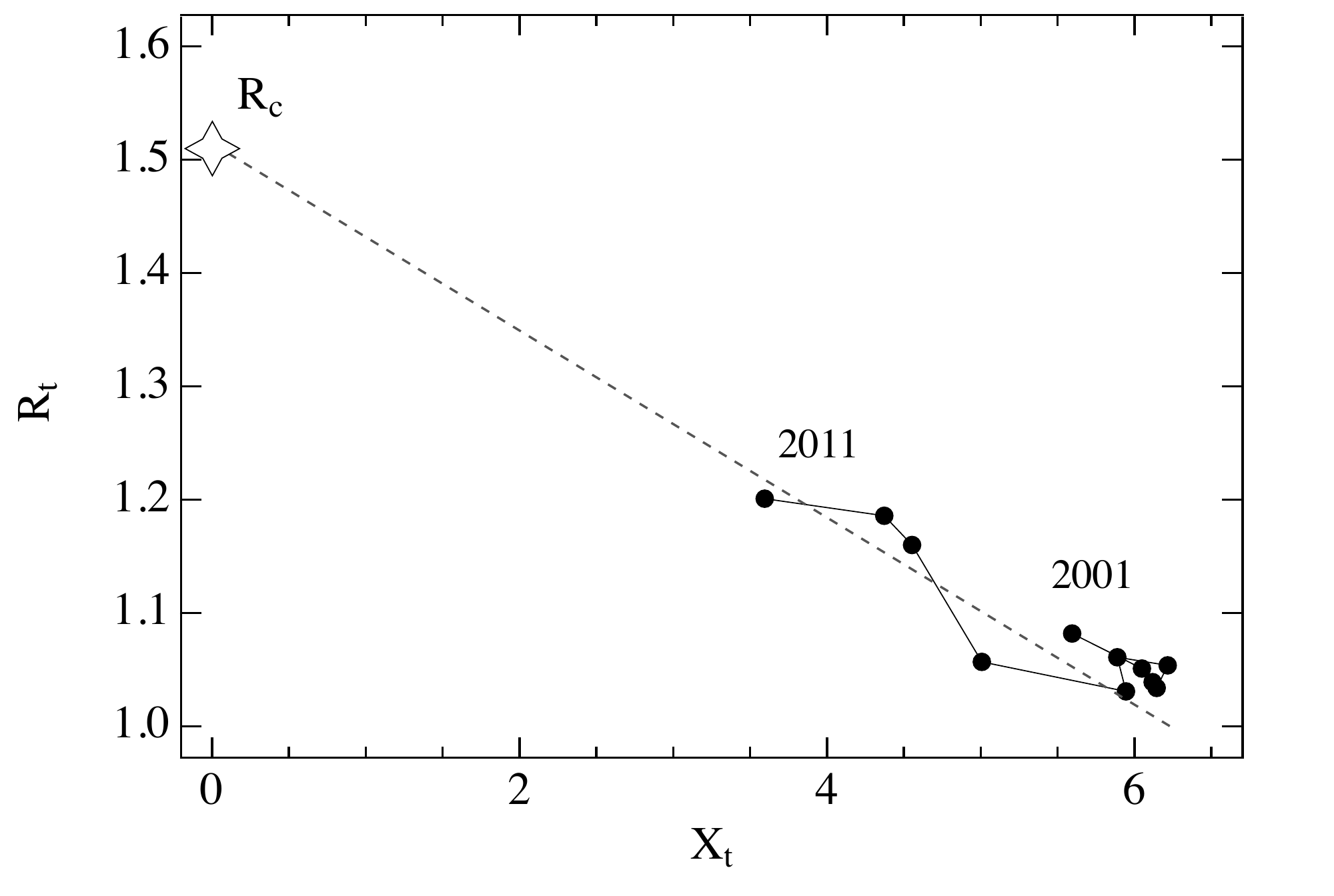}}
\href{http://www.necsi.edu/research/economics/bondprices/italy2.pdf}{\includegraphics[width=0.49\linewidth]{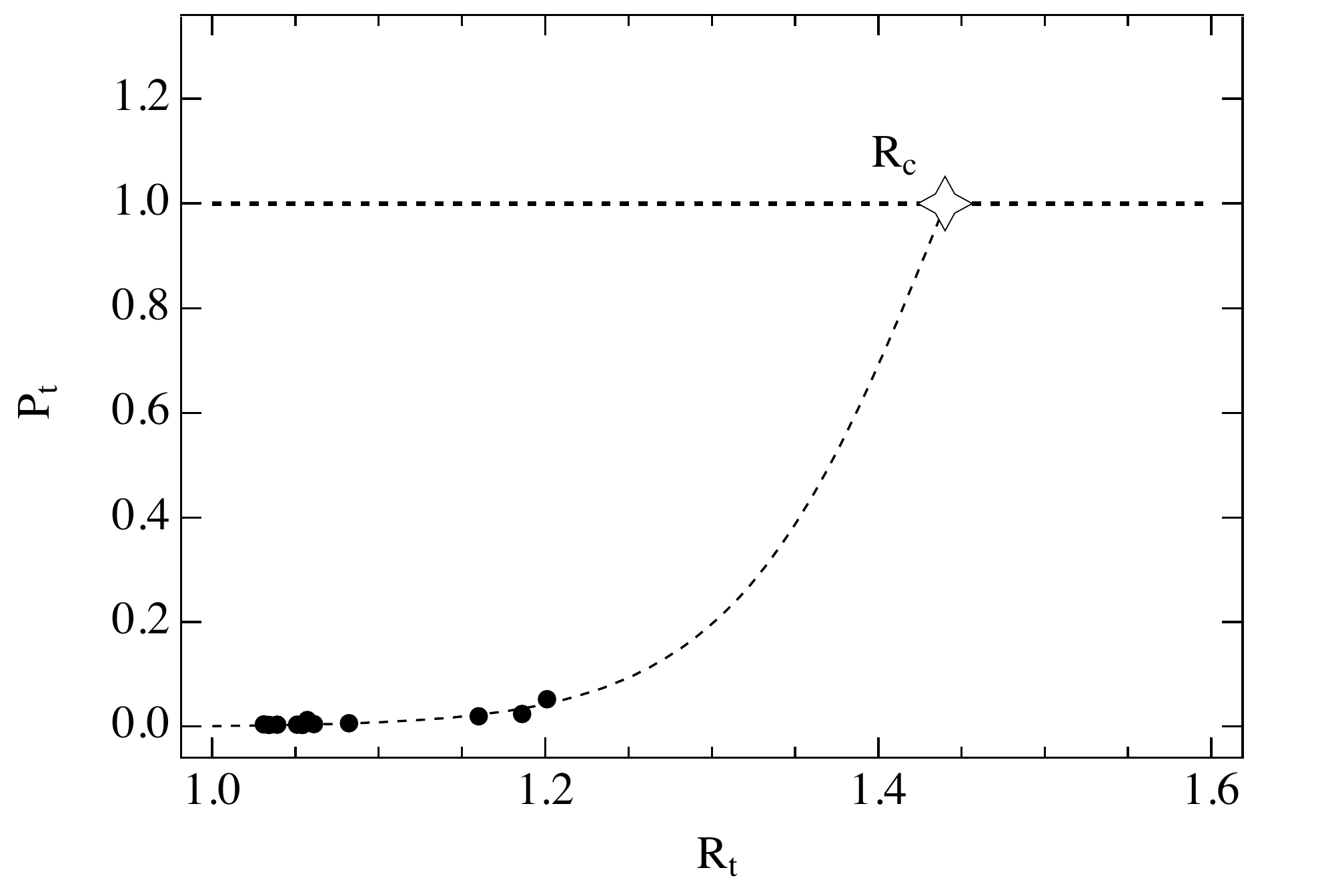}}
\href{http://www.necsi.edu/research/economics/bondprices/italy3.pdf}{\includegraphics[width=0.49\linewidth]{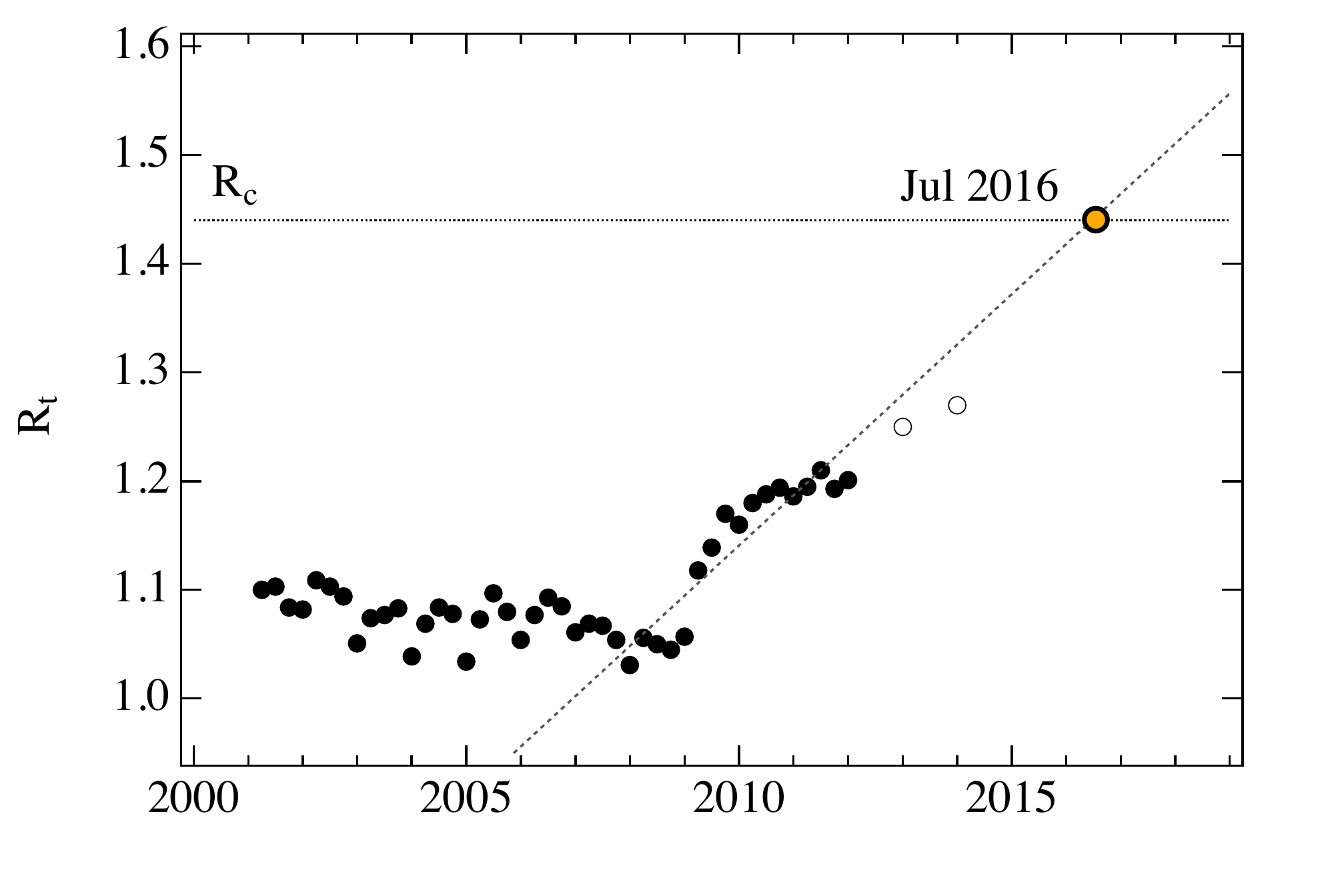}}
\caption{\textbf{Model fits and projections for Italy} - \emph{Top left panel}: Italy's debt ratio \cite{eurostat} as a function of the default distance parameter $X_t$ (Eq. \ref{eq:linear2}) from 2001 to 2011 inclusive. $X_t$ reflects long-term interest rates, i.e. secondary market yields of 10-year bonds \cite{oecd}. Data (solid line) can be fitted with Eq. \ref{eq:linear2} (dashed line) from 2007 to 2011 to obtain the two model parameters, $R_c=1.51\pm0.1$ and $\eta = 0.08\pm0.02$. \emph{Top right panel}: Probability of default as a function of Italy's debt ratio. $P_t$ is calculated from Eq. \ref{eq:pdefault}, setting the recovery rate to $\rho =0.5$ and considering German interest rates as the risk-free interest rates. Data (dots) can be fitted with Eq. \ref{eq:final_p} (dashed line). This allows us to extract the two model parameters, the critical debt ratio $R_c=1.44 \pm 0.04$ and the heterogeneity parameter $\eta = 0.06\pm0.01$. \emph{Bottom panel}: Italy's debt ratio (solid dots) as a function of time. The debt trajectory, projected with a linear regression (dashed line), intersects the critical debt ratio ($R_c=1.44$) in Jul 2016. Hollow circles are projected debt ratios for the end of 2012 and 2013 if austerity measures are met (calculated from \cite{irishtimes2012,Donadio2012,Brunsden2012}).}
\end{figure}

\end{document}